\documentclass[default,iicol]{sn-jnl}% Default with double column layout

\newcommand{\argmax}{\operatornamewithlimits{argmax}}
\usepackage{subcaption}
\usepackage{array}
\usepackage{bbold}
\usepackage{url}
\newcolumntype{H}{>{\setbox0=\hbox\bgroup}c<{\egroup}@{}}

\newcommand*{\defeq}{\mathrel{\vcenter{\baselineskip0.5ex              \lineskiplimit0pt
    \hbox{\scriptsize.}\hbox{\scriptsize.}}}%
     =}
     
\jyear{2021}%

\theoremstyle{thmstyleone}%
\theoremstyle{thmstyletwo}%

\theoremstyle{thmstylethree}%

\begin{document}

\title[ ]{Efficient and accurate inference for mixtures of Mallows models with Spearman distance}


\author[1]{\fnm{Marta} \sur{Crispino}}\email{marta.crispino@bancaditalia.it}
\equalcont{These authors contributed equally to this work.}

\author[2]{\fnm{Cristina} \sur{Mollica}}\email{cristina.mollica@uniroma1.it}
\equalcont{These authors contributed equally to this work.}

\author[1]{\fnm{Valerio} \sur{Astuti}}\email{valerio.astuti@bancaditalia.it}

\author[2]{\fnm{Luca} \sur{Tardella}}\email{luca.tardella@uniroma1.it}

\affil[1]{ 
\orgdiv{DG of Economics, Statistics and Research}, \orgname{Bank of Italy}, Rome, Italy}

\affil[2]{Department of Statistical Sciences, Sapienza University of Rome, Italy}


\abstract{The Mallows model occupies a central role in parametric modelling of ranking data to learn preferences of a population of judges. Despite the wide range of metrics for rankings that can be considered in the model specification, the choice is typically limited to the Kendall, Cayley or Hamming distances, due to the closed-form expression of the related model normalizing constant.  
This work instead focuses on the Mallows model with Spearman distance.%, also known as $\theta$-model.

An efficient and accurate EM algorithm for estimating finite mixtures of Mallows models with  Spearman distance is developed, by relying on a twofold data augmentation strategy aimed at i) enlarging the applicability of Mallows models to samples drawn from heterogeneous populations; ii) dealing with partial rankings affected by diverse forms of censoring. Additionally, a novel approximation of the model normalizing constant is introduced to support the challenging model-based clustering of rankings with a large number of items. The inferential ability of the EM scheme and the effectiveness of the approximation are assessed by extensive simulation studies. Finally, we show that the application to three real-world datasets endorses our proposals also in the comparison with competing mixtures of ranking models. %200 or fewer words
}




\maketitle

\section{Introduction}

Ranking data arise when a sample of $N$ people is presented a finite set of $n$ alternatives, called \textit{items}, and is asked to rank them according to a certain criterion, such as personal preferences or attitudes. Thus, a ranking is the result of a comparative judgment on a set of competing alternatives expressed in the form of order relation. 

The increasing interest in ranked data analysis is motivated by several research contexts, among which marketing and political surveys, where items could be consumer goods, political candidates or goals, but also by psychological and behavioral studies consisting, for instance, in ordering of words/topics according to the perceived association with a reference subject \citep{gormley2008exploring,Caron2014}. 
%However, ranking data are not limited to the inquiry of preferences or attitudes of a target population. 
Another typical context for ranking data
%that naturally gives rise to ranking data 
is sports and, more generally, competitions. Some examples are horse or car races, soccer and basket championships, where the rankings are determined by the relative ability of the competing players or teams 
%\citep{Henery-Royal,Benter,Henderson}.
\citep{Henery-Royal}.

However, ranking data are not necessarily the result of a human comparative choice where, typically, only a limited number of options can be submitted to the judgment of the rankers. %In fact, 
The need of handling large ordered lists of alternatives can be motivated, for example, with the analysis of the output obtained from search algorithms. This can include the results of web-search engines, as well as of programs for matching biological sequences or specific traits with a reference database, that return a list of items ordered according to the degree of agreement with the input information \citep{DeConde2006}. %\cm{add references such as Liu}. 
A further field of application concerns the widespread digital platforms of on demand services, such as movies, songs and any kind of product available from the online marketplaces. In this context, various forms of feedback from the customers (e.g. clicks, ratings, reviews and purchases) are elaborated from the e-commerce companies to capture customers' preferences and supply them with a personalized ordered list of recommended items
%products to the users depending on the features and frequencies of their previous choices 
%\cm{add references}
\citep{sylvia}. All these applications are characterized by the challenging task of dealing with a large value $n$ of items which are often not completely ranked, 
%that is, only for some of these numerous alternatives the position is exactly known, 
giving rise to \textit{partial rankings}.

%Let $\boldsymbol{r}=(r_1,\dots,r_n)$ be a ranking of the $n$ items, with the component $r_i$ indicating the rank assigned to item $i$. We adopt the usual convention that $r_i<r_i'$ means that item $i$ is preferred to $i'$ (the lower the rank is, the more preferred is the item). Both items and ranks are identified with the set $\{1,\dots,n\}$, implying that a generic observation $\boldsymbol{r}$ is a permutation of the first $n$ integers and, thus, a point in the finite discrete permutation space $\mathcal{P}_n$.% with size equal to $n!$. 

In the wide literature on probability distributions for
%random permutations, 
ranking data, see \cite{Marden1995} and \cite{AlvoYu2014} for extensive reviews, the class of \textit{Mallows models} (MM) became very popular for its long history, both as a successful approach for preference learning and as a starting point for effective  methodological advances. 
%The paternity of this parametric class is originally attributed to \cite{Mallows1957}, who introduced it by focusing on the Kendall and Spearman metrics for the space $\mathcal{P}_n$. However, the formalization of the Mallows approach as a specific method for model building based on the use of several possible metrics for the permutation space is due to \cite{Diaconis1988}.
Formally, the MM assumes the existence of a unique modal ranking %$\boldsymbol\rho\in\mathcal{P}_n$ 
representing the common comparative judgment on the items, usually referred to as \textit{consensus ranking}. Moreover, the variability of the population of rankers, that is, the agreement with the consensus ranking, is calibrated by a \textit{concentration parameter}. %$\theta$. 
A third key ingredient of the MM
%mathematical form of the 
 is the distance between rankings, since
%postulates that 
the probability of a generic 
%ranking decreases exponentially as the distance 
sequence depends on its distance from the consensus; as a consequence the MM is also referred to as \textit{distance-based model} (DBM).
%$\boldsymbol\rho$ increases. 

Even if these models are intuitive and attractive, especially for the fact that they represent exponential families for random permutations \citep{Diaconis1988}, postulating a MM as a generative process of the observed rankings does not prevent noticeable practical issues. First, one of its peculiarities is the mixed-type parameter space, combining the continuous concentration parameter
%$\theta$ 
and the discrete consensus ranking. The latter, in fact,
%$\boldsymbol{\rho}$. 
takes values in the space $\mathcal{P}_n$ of the $n!$ permutations of the first $n$ integers, whose adequate exploration, needed for making inference on the model, is computationally demanding
%or even infeasible 
and challenging as the number of items increases. Moreover, inference is difficult to perform in those cases when the model normalizing constant is intractable.
%In fact, the rapidly-increasing size $n!$ makes the exhaustive enumeration of the sequences in $\mathcal{P}_n$ feasible only for few values of $n$. 
Second, the estimation task is further complicated in the presence of partial observed rankings and of heterogeneity in the sample. Finally, some criticism was raised about the low flexibility of the MM \citep{mollica14}, which can limit its adequacy in describing many real-world cases, especially when the sample size is high. 

For the above reasons, in the last decades various directions to generalize the MM and improve its fitting performance have been proposed, among which the mixture approach plays a leading role. Nevertheless, a review of the existing MM extensions reveals that the version of the MM with Spearman distance (MMS henceforth) received very little attention, even if this metric induces a probability distribution on the permutation space that parallels the Gaussian model and, similarly, enjoys convenient analytical simplifications. The main reason for this lack of attention lies in the fact that the MMS is characterized by an intractable normalizing constant, while the MM with other distances (such as the Kendall and the Cayley) enjoys favourable closed-form expressions
%of the normalizing constant 
\citep{Fligner1986, IrurozkiThesis}.

This work focuses on the MMS, originally introduced by \cite{Mallows1957} under the name $\theta$-model, and its extension into the finite mixture framework %within the frequentist inferential domain 
for the analysis of full and partial rankings. Specifically, 
%by starting from the result provided by \citep{Marden1995}, proving that the MMS admits a convenient closed-form solution for the critical maximum likelihood estimation (MLE) of the modal sequence, %$\boldsymbol{\rho}$, 
we exploit the crucial result that, unlike other DBMs, the MMS admits an 
%shares the 
analytical solution for the critical estimation of the consensus ranking
%under the uniform (null) model specification, provided in \citep{Marden1995}, 
%and build on this crucial result
%, pointed out in \citep{FeiginCohen}, 
to generalize the MMS within the finite mixture approach 
and perform model-based clustering
%for making inference on this model from 
of arbitrary forms of partial rankings
%an arbitrary form of censoring 
from the maximum likelihood estimation (MLE) perspective. % \mc{io non la metterei giù così: lui ha il teorema, ma non lo applica al caso di MM. controlliamo però.}. 
Moreover, we propose an original approximation of the 
MMS normalizing constant,
%distribution of the Spearman distance, 
aiming at effectively addressing the computational issues arising when inference is conducted on rankings of a large number $n$ of items. This approximation, which may be interesting in its own right (for instance for sampling procedures), is then compared with existing approximating approaches.
%\mc{in fact, we do propose an approximation that characterizes the distribution of the Spearman distance, with the aim at effectively addressing the computational issues arising when inference is conducted on rankings of large number $n$ of items.(for moderate values of $n$, while the asymptotic is given by \citep{kendall1970rank}). This approximation, which may be interesting in its own right (e.g. for sampling procedures) is then used to approximate the partition function of the MMS and compared with existing approximating approaches, with the aim at effectively addressing the computational issues arising when inference is conducted on rankings of large number $n$ of items.} 
Inference is addressed with a new Expectation-Maximization (EM) algorithm which is much more accurate and efficient than existing versions, and allows to successfully investigate the group structure of large samples of rankings with a high number of alternatives. The merits of our inferential approach are illustrated with applications to simulated and real-world ranking data. 

The outline of the paper is the following. In Section \ref{sec:model}, after reviewing the MM and its main generalizations, we focus on the inference on the MMS. We then move to detailing the finite mixture framework, also sketching the EM algorithm, and finally we explain the data augmentation strategy to deal with heterogeneous partial rankings. In Section \ref{ssec:approx}, we present the new approximation of the model normalizing constant and compare it with available alternatives. An extensive simulation study is performed in Section \ref{sec:simu}, by considering a wide range of scenarios in terms of sample size, number of items,
%for the homogeneous and 
heterogeneity of the sampling distribution, as well as different patterns of data missingness. Section \ref{sec:applications} deals with real-world data illustrations, then the paper ends with concluding remarks in Section \ref{sec:conc}.

\section{Model and inference}\label{sec:model}
%In this section we briefly review the general MM,
%and then focus on inference in the MM with Spearman distance (MMS).
%We then deal with the MMS mixture model for full rankings
%and introduce its brand new extension for partial rankings.
%Finally, we briefly comment on the criterion used for the automatic selection of the number of mixture components.

\subsection{The Mallows model: a brief review}\label{ssec:review}

Let $\boldsymbol{r}=(r_1,\dots,r_n)$ be a ranking of $n$ items with generic entry $r_i$ indicating the rank assigned to item $i$.
We adopt the usual convention 
that $r_i<r_{i'}$ means that item $i$ is preferred to item $i'$ (the lower the rank is, the more preferred the item is). 
Both items and ranks are identified with the set $\{1,\dots,n\}$, implying that a generic observation $\boldsymbol{r}$ is a permutation of the first $n$ integers and, thus, a point in the finite discrete space $\mathcal{P}_n$.% with size equal to $n!$.

The MM is a well-established parametric class of ranking distributions whose paternity is attributed to \cite{Mallows1957}. The probability of observing the ranking $\boldsymbol{r}$ under the MM is
 \begin{equation}
 \label{eq:MM}
%   \mathbb{P}(\boldsymbol R =\boldsymbol{r}\,\vert\boldsymbol{\rho},\theta)\defeq
\mathbb{P}(\boldsymbol{r}\,\vert \boldsymbol{\rho},\theta)
=\frac{e^{-\theta\, d(\boldsymbol{r},\boldsymbol\rho)}}{Z(\theta,\boldsymbol\rho)}
%=\frac{e^{-2\theta\, \left(c_n-\boldsymbol{\rho}^T\boldsymbol{r}\right)}}{Z(\theta,\boldsymbol\rho)},
\qquad\boldsymbol r\in\mathcal{P}_n,
 \end{equation}
where $\boldsymbol\rho\in\mathcal{P}_n$ is the consensus ranking, $\theta\in\mathbb{R}^+_0$ is the concentration parameter, $d(\cdot,\cdot)$ is a distance over $\mathcal{P}_n$ and $Z(\theta,\boldsymbol\rho)=\sum_{\boldsymbol{r} \in \mathcal{P}_{n}} e^{-\theta\, d(\boldsymbol{r},\boldsymbol\rho)}$ is the normalizing constant or \textit{partition function}. From \eqref{eq:MM} it follows that, under the MM assumption, the probability of observing a ranking decreases exponentially as its distance from the consensus $\boldsymbol\rho$ increases. %\mc{forse qui aggiungere un discorso sulla difficoltà di avere a che fare con Z, in generale, in modo da inquadrare subito il problema. }

In the attempt to enlarge the flexibility of the MM and improve the goodness-of-fit, various generalizations appeared in the literature. \cite{Fligner1986} proposed the \textit{Generalized Mallows model} (GMM), which considers a $(n-1)$-dimensional vector of concentration parameters, each affecting a particular position in the ranking. The GMM, however, can be defined only when the distance is decomposable into the sum of independent components associated to the each single stage of the ranking process, such as for the Kendall, Cayley and Hamming metrics. %, a property shared by the Kendall and the Cayley metrics only.
\cite{MeilaBao2010} extended the GMM to infinite rankings with the \textit{Infinite Generalized Mallows model}, to handle those cases where the number of items is very high, or potentially not completely known, such as in the output of search engines. Another proposal to enhance the flexibility of the MM is  the \textit{weighted distance-based model} (WDBM) by \cite{lee2010distance}, which amounts to use weighted versions of the commonly used distances between rankings, including the Spearman one.

In order to explore the unobserved heterogeneity in the sample, the aforementioned extensions were considered also in the finite mixture modelling approach. For example, \cite{MurphyMartin2003} described the general EM algorithm to derive the MLE of MM mixtures for an arbitrary metric; \cite{lee2012mixtures} detailed the same inferential method for the mixture of WDBMs. Differently from the one illustrated by \cite{MurphyMartin2003}, \cite{BusseEtal2007} introduce an EM algorithm for the MM mixtures, working with Kendall distance only but handling both complete and partial rankings of a larger number of items. 

However, all the above contributions do not specifically focus on the MMS and, if they handle the Spearman distance, they only work in practice with few items (less than $n=10$, say). 
This work, instead, is devoted to the MMS and its mixture extension, with the aim of developing a learning strategy able to handle also rankings of a large number of items.
%, a probability  distribution  on  the permutation space that parallels the Gaussian model.
%The next sections are therefore devoted to the MMS and its finite mixture extension. 

\subsection{Likelihood and inference for the MMS}
\label{ss:MLEhomo}
The Spearman distance between two permutations $\boldsymbol{r},\boldsymbol\rho\in \mathcal{P}_n$ is defined as follows
\begin{equation}
\label{eq:spearman}
d(\boldsymbol{r},\boldsymbol\rho)=\sum_{i=1}^n(r_i-\rho_i)^2.
\end{equation}
It is an unnormalized version of the Spearman rank correlation, used to measure the statistical correlation between the ranks of two variables but, when rankings are considered as vectors in $\mathbb R^n$, it is simply the squared Euclidean distance, or $L_2$-norm. By developing the square in \eqref{eq:spearman} and setting $c_n=n(n+1)(2n+1)/6$ and the scalar product $\boldsymbol{\rho}^T\boldsymbol{r}=\sum_{i=1}^n\rho_ir_i$, where the symbol $^T$ denotes the transposition (row vector), one has
$d(\boldsymbol{r},\boldsymbol{\rho})=2\left(c_n-\boldsymbol{\rho}^T\boldsymbol{r}\right)
$. Hence, the MMS can be written as
%assumes that the probability of observing the generic ranking $\boldsymbol r\in\mathcal{P}_n$ is
 \begin{equation}
 \label{eq:ProbMassMallows}
%   \mathbb{P}(\boldsymbol R =\boldsymbol{r}\,\vert\boldsymbol{\rho},\theta)\defeq
\mathbb{P}(\boldsymbol{r}\,\vert\boldsymbol{\rho},\theta)
%=\frac{e^{-\theta\, d_S(\boldsymbol{r},\boldsymbol\rho)}}{Z(\theta,\boldsymbol\rho)}
%=\frac{e^{-2\theta\, \left(c_n-\boldsymbol{\rho}^T\boldsymbol{r}\right)}}{Z(\theta,\boldsymbol\rho)},
=\frac{e^{-2\theta\, \left(c_n-\boldsymbol{\rho}^T\boldsymbol{r}\right)}}{Z(\theta)},
 \end{equation}
where the normalizing constant $Z(\theta)=\sum_{\boldsymbol{r} \in \mathcal{P}_{n}} e^{-2\theta\, \left(c_n-\boldsymbol{e}^T\boldsymbol{r}\right)}$ with $\boldsymbol e=(1, 2, ..., n)$
%denoting the identity permutation, 
does not depend on $\boldsymbol\rho$ because of the right-invariance property of the Spearman distance \citep{Diaconis1988}.
Note that the number of terms in the sum grows with $n!$, which makes direct calculation of $Z(\theta)$ unfeasible for all but very small values of $n$, typically
%it is possible to exactly compute it 
for at most $n=11$.
%is the normalizing constant (or \textit{partition function}), $\boldsymbol{\rho}\!\in\!\mathcal{P}_n$ is a location parameter representing the shared consensus ranking and $\theta\!\geq\!0$ is a scale parameter describing the concentration of the distribution around the global consensus.  Thanks to the property of right-invariance of the Spearman distance \citep{Diaconis1988}, meaning that the metric does not depend on the arbitrary relabelling of the items, the partition function simplifies into a function of the concentration parameter only, given by
%$$Z(\theta,\boldsymbol\rho)=Z(\theta,\mathbf{e})=Z(\theta)=\sum_{\boldsymbol{r} \in \mathcal{P}_{n}} e^{-\theta \, \sum_{i=1}^n(r_i-i)^2},$$ 
%with $\mathbf{e}=(1,2,3,\ldots,n)$ indicating the identity permutation. 

Let $\underline{\boldsymbol{r}}=(\boldsymbol{r}_1,\dots,\boldsymbol{r}_N)$ be a random sample of size $N$ drawn from the MMS and $N_l$ be the frequency of the $l$-th distinct observed ranking $\boldsymbol{r}_l$ for $l=1,\dots,L$, such that $\sum_{l=1}^L N_l=N$. The observed-data log-likelihood of the MMS is 
\begin{equation}
\label{likelihood}
\begin{split}
%p(\boldsymbol{r}_1,\dots,\boldsymbol{r}_N\vert\boldsymbol{\rho},\theta)=
\ell(\boldsymbol{\rho},\theta\vert\underline{\boldsymbol{r}})
%&=-N\log{Z(\theta)}-\theta\sum_{j=1}^N\sum_{i=1}^n(r_{ji}-\rho_i)^2\\
=-N\left[\log{Z(\theta)}+2\theta\left(c_n-\boldsymbol{\rho}^T{\boldsymbol{\bar r}}\right)\right],
\end{split}
\end{equation}
where ${\boldsymbol{\bar r}}=(\bar{r}_1,\ldots,\bar{r}_n)$ is the sample mean rank vector whose $i$-th entry is $\bar{r}_i=\frac{1}{N}\sum_{l=1}^LN_lr_{li}$. 
When considering the
%squared Euclidean distance as 
metric \eqref{eq:spearman} in the MM, it is not surprising that the likelihood of the MMS resembles that of an $n$-variate normal, except for its finite support. In other words, the MMS is the restriction of the $n$-dimensional Gaussian on $\mathcal{P}_n$ \citep{crispino2022informative}. 
It follows that, if $\boldsymbol{\bar r}\in\mathcal P_n$, then the MLE $\hat{\boldsymbol\rho}$ of the consensus ranking simply coincides with the sample mean rank vector. However, in general $\boldsymbol{\bar r}\notin\mathcal P_n$, so a further consideration is required in order to solve the optimization problem. Specifically, the log-likelihood expression \eqref{likelihood} implies that
%, for $\theta>0$, 
the MLE of the consensus is
\begin{equation}
\label{optimum}
\hat{\boldsymbol\rho}
%=\argmin_{\boldsymbol{\rho}\in\mathcal{P}_n}\sum_{j=1}^Nd_S(\boldsymbol{r}_j,\boldsymbol\rho)
=\argmax_{\boldsymbol{\rho}\in\mathcal{P}_n}\boldsymbol\rho^T\boldsymbol{\bar r}.
\end{equation}
We observe that the optimization problem \eqref{optimum} is equal to the one discussed in \cite{Marden1995} for estimating the central ranking under uniformity of the random permutations, that the author proved to admit the closed-form solution
%given by
\begin{equation}\label{eq:borda}
    \hat{\boldsymbol{\rho}}=(\hat\rho_1,\ldots,\hat\rho_i,\ldots,\hat\rho_n)\,\,\,\text{with}\,\,\,\hat\rho_i=\text{rank}(\bar{r}_i)%\text{ in }\{\bar{r}_1,\dots,\bar{r}_n\}
\end{equation}
%when all the mean ranks $\bar{r}_i$ are distinct 
(Theorem 2.2, page 29). 
%\mc{aggiustare} 
This implies that the simple and fast Borda rank aggregation method, which amounts to order the items according to their sample average rank, 
%(over the $N$ observed ones) 
%\citep{rankAggreg},
%As a consequence, the ranking \eqref{eq:borda} coincides with the ranking arising from the Borda count.} %provides 
%the exact solution of \eqref{optimum} also for the MMS 
coincides with the MLE of the consensus ranking of the MMS. 
%\citep[note that a version of this result was also pointed out in][]{FeiginCohen}. 
%In practice, however, in recent years the use of Borda method, was only recommended for the MM with the Kendall distance (MMK) either as approximation of the MLE of $\boldsymbol\rho$ \citep{Fligner:Verducci-American,caragiannis2013noisy,Ali2012,IrurozkiThesis} or for initializing the local search \citep{irurozki2016permallows, rankdist}.  
%The Borda count, in fact, is an unbiased estimator of the Kemeny ranking which is the MLE of the MM with the Kendall distance.} 
Even if this theoretical result, pointed out for the MMS in \cite{FeiginCohen}, efficiently solves the critical step of estimating the discrete parameter $\boldsymbol\rho$ regardless of the number $n$ of compared items and the sample size $N$,
%to be ranked, 
to the best of our knowledge it has been completely disregarded in those few works dealing with the MMS and extensions thereof. In fact, the optimization \eqref{optimum} has been traditionally addressed
%in the same manner as for those distances for which 
as if a closed-form solution did not exist, that is, with computationally intensive approaches, such as global or local search methods whose effectiveness and applicability strongly depend on the values of $n$ and $N$ \citep[see for example][]{MurphyMartin2003,BusseEtal2007}. These search methods require the exploration of a discrete space with a fast-growing dimension and, hence, are inefficient both in terms of execution time and convergence achievement as $n$ and $N$ increase, becoming soon infeasible  \citep[typically, NP-hard as proven in][]{bartholdi1989computational}.
The omission of this important result 
%forand the inference for the discrete MMS parameter 
is also testified by the lack of specific and effective software for performing inference on the MMS, that could take advantage of the related computational simplifications. 
In fact, the use of the Borda method has been so far mainly recommended for the MM with Kendall distance, either as approximation of the MLE of $\boldsymbol\rho$ \citep{Ali2012} or for initializing the local search \citep{irurozki2016permallows}.

Concerning the estimation of the concentration parameter $\theta$, one needs to 
%employ 
maximize the profile log-likelihood $\ell(\hat{\boldsymbol{\rho}},\theta\vert\underline{\boldsymbol{r}})$ to get %the MLE as follows 
%Moreover, one can get 
%$$\hat\theta = \underset{\theta\in\mathbb{R}^+_0}{\argmax}\ \ell(\hat{\boldsymbol{\rho}},\theta\vert\underline{\boldsymbol{r}}),$$
%as the solution
%recalling equality~\eqref{e:zmrel}, to $\hat\theta$
%yielding 
$\hat\theta$ as the solution of the following equation
\begin{equation}
\label{e:thetaest}
-Z'(\theta)/Z(\theta)
=2(c_n-\hat{\boldsymbol{\rho}}^T{\boldsymbol{\bar r}}).
%\bar{d}_K=\dfrac{M_{D_K}'(-\theta)}{M_{D_K}(-\theta)}
%\Longleftrightarrow
%\bar{d}_K=\dfrac{d}{dt}\log M_{D_K}(t)_{\big\rvert_{t=-\theta}},
\end{equation}
%
%where $-Z'(\theta)/Z(\theta)=\text{E}_\theta(D)$, that is  
As noticed by \cite{Fligner1986}, $\hat\theta$ is the value equating the expected Spearman distance $\text{E}_{\theta}[D]$ under the MMS, that is, the left hand side of \eqref{e:thetaest}, to the sample average Spearman distance $\bar{d}=\frac{1}{N}\sum_{l=1}^LN_ld(\boldsymbol{r}_l,\hat{\boldsymbol{\rho}})$, corresponding to the right-hand side of \eqref{e:thetaest}.
%, to , corresponding to the left-hand side of \eqref{e:thetaest}. 
Actually, $\text{E}_\theta[D]$ is a strictly decreasing function of $\theta$, but the root of the above equation is not available in closed-form. % \mc{because of the intractability of the normalizing constant $Z(\theta)$}\cm{in realtà pure con le altre distanze questa radice viene trovata numericamente, dunque non farei questa precisazione}. 
So, the MLE can be found numerically, for instance, via a Newton-Raphson algorithm \citep{Marden1995}, although the calculations are demanding for all but small values of $n$, because of the intractability of the model normalizing constant. 

Inference in the presence of intractable normalizing constants has been tackled in the literature either with algorithms aimed at finding consistent estimators for the unknown parameters \citep[see for example][]{murray2012mcmc,Andrieu2009}, either with off line approximations of the normalizing constant \citep{vitelli18,XuAlvoYu2018}. 
The first stream of methods is unfeasible in case of the MMS, because it
%is 
typically
%necessary to be able to simulate directly 
requires the direct simulation from the model likelihood \citep{vitelli18}. We here propose to solve the inferential issue by introducing a new analytical approximation of $Z(\theta)$ (see Section \ref{ssec:approx}), whose performance is compared with alternative ones.

%In Section \ref{ssec:approx}, we introduce an approximation to address this issue. 

\subsection{Likelihood and inference for the MMS mixture model}
\label{ss:EM}

Let $\boldsymbol{z}_l=(z_{l1},\dots,z_{lG})$ be the latent variable  collecting the binary group membership indicators, that is, $z_{lg}=1$ when the $l$-th observation belongs to the $g$-th group and $z_{lg}=0$ otherwise.
%
%\begin{equation*}
%z_{lg}=
%\begin{cases}
%1\qquad\text{if the $l$-th unit(s) belongs to the $g$-th group}, \\
%0\qquad\text{otherwise}.
%\end{cases}
%\end{equation*}
%
The complete-data log-likelihood for the $G$-components mixture of \mbox{MMS is}
\begin{equation*}
\label{e:mDB}
\begin{split}
\ell_c(\boldsymbol{\rho},\boldsymbol{\theta},\boldsymbol{\omega},\underline{\boldsymbol{z}} \vert \underline{\boldsymbol{r}}) %;\underline{\pi}
%&=\log \mathbb{P}(\underline{\pi},\underline{z}\vert\usigma,\ulambda,\uomega)
%=\log\prod_{j=1}^N\mathbb{P}(\boldsymbol{r}_j,\underline{z}_s\vert\usigma,\ulambda,\uomega)\\
%&=\log\prod_{j=1}^N\mathbb{P}(\boldsymbol{r}_j\vert\underline{z}_s,\usigma,\ulambda)\mathbb{P}(\underline{z}_s\vert\uomega)\\
%&=\sum_{j=1}^N\sum_{g=1}^G\log\biggl({\omega_g\dfrac{e^{-\theta_gd_K(\boldsymbol{r}_j,\sigma_g)}}{Z(\theta_g)}\biggr)}^{z_{jg}}\\
%=\sum_{j=1}^N\sum_{g=1}^Gz_{jg}\bigl(\log\omega_g- 2\theta_g\left(c_n-{\boldsymbol{\rho}_g}'\boldsymbol{r}_j\right)-\log Z(\theta_g)\bigr), %Marta:rimosso hat su rho_g
=&\sum_{l=1}^L\sum_{g=1}^GN_lz_{lg}\big[\log\omega_g+\\
&-2\theta_g\left(c_n-{\boldsymbol{\rho}^T_g}\boldsymbol{r}_l\right)-\log Z(\theta_g)\big], %Marta:rimosso 
\end{split}
\end{equation*}
where $\boldsymbol{\omega}=(\omega_1,\dots,\omega_G)$ and $\boldsymbol{\theta}=(\theta_1,\dots,\theta_G)$ are, respectively, the 
%group membership probabilities 
mixture weights and the component-specific concentration parameters, whereas $\boldsymbol{\rho}$ is a $G\times n$ matrix, whose rows indicate the consensus rankings of the mixture components. In order to perform MLE  in the presence of latent variables, we employ the EM algorithm \citep{Demp:Lai:Rub}. 
%Note that \citep{MurphyMartin2003} first provided the EM algorithm for the MLE of MM mixtures with a generic metric $d(\cdot,\cdot)$, including the Spearman. However, their estimation procedure is not specialized for each considered metric implying that, in the case of Spearman distance, it completely disregards the closed-form solution of the M-step for $\boldsymbol{\rho}$ whereas, for the decomposable metrics, the convenient expression of the normalizing constant is totally ignored. Actually, their work was limited to applications on full rankings of only $n=5$ items, for which the permutation space can be explored exhaustively and, even with a time-consuming inferential process, the MLE is more likely to be achieved. Instead, our implementation of the EM is completely specialized for the Spearman distance, thus exploiting the computational simplifications explained in the previous sections. 
In particular, at iteration $(t+1)$, the proposed EM scheme consists in the following steps:
\vspace{0.3cm}
\hspace{-0.3cm}\fbox{%
\begin{minipage}{2.8in}
\scriptsize
\begin{description}
\item[\textbf{E-step:}] 
for $l=1,\dots,L$ and $g=1,\dots,G$,
\vspace{-0.1cm}
\begin{itemize}
\scriptsize
\item[-] compute the posterior membership probabilities
$$\hat z_{lg}^{(t+1)}=\dfrac{\hat{\omega}_g^{(t)}\mathbb{P}\left(\boldsymbol{r}_l\big\vert\hat{\boldsymbol{\rho}}_g^{(t)},\hat{\theta}_g^{(t)}\right)}{\sum_{g'=1}^G\hat{\omega}_{g'}^{(t)}\mathbb{P}\left(\boldsymbol{r}_l\big\vert\hat{\boldsymbol{\rho}}_{g'}^{(t)},\hat{\theta}_{g'}^{(t)}\right)}.$$
\end{itemize}
\item[\textbf{M-step:}] %\phantom{mmm}]
for $g=1,\dots,G$,
\vspace{-0.1cm}
\begin{itemize}
\scriptsize
    \item[-] compute the mixing weights $$\hat\omega_g^{(t+1)}=\frac{\hat{N}_g^{(t+1)}}{N}$$
    where $\hat{N}_g^{(t+1)}=\sum_{l=1}^LN_l\hat z_{lg}^{(t+1)}$
    is the number of sample units allocated in the $g$-th component.\\
    \item[-] compute the consensus rankings $$\hat{\boldsymbol{\rho}}_g^{(t+1)} = \left(\hat\rho_{g1}^{(t+1)},\ldots,\hat\rho_{gi}^{(t+1)},\ldots,\hat\rho_{gn}^{(t+1)}\right),$$ with  $\hat{\rho}_{gi}^{(t+1)}=\text{rank}\left(\bar{r}_{gi}^{(t+1)}\right)$  and 
$\bar{r}_{gi}^{(t+1)}=\sum_{l=1}^LN_l\hat z_{lg}^{(t+1)}r_{li}/\hat{N}_g^{(t+1)}$.\\
%corresponding to the average rank of item $i$ in cluster $g$;
\item[-] determine the concentration parameter estimates $\hat{\theta}_g^{(t+1)}$ as the solution of 
$$\text{E}_{\theta_g}[D]
=2\left(c_n-\boldsymbol{\rho}_g^{T(t+1)}{\boldsymbol{\bar r}}_g^{(t+1)}\right).$$
\end{itemize}
\end{description}
\end{minipage}
}
\vspace{0.3cm}
%\hspace{1cm}

Note that \cite{MurphyMartin2003} first provided an EM algorithm for the MLE of MM mixtures with a generic metric $d(\cdot,\cdot)$, including the Spearman. However, their estimation procedure is not specialized for each considered metric implying that, in the case of Spearman distance, it completely disregards the closed-form solution of the M-step for $\boldsymbol{\rho}$ whereas, for the decomposable metrics, the convenient expression of the normalizing constant is totally ignored. Actually, the applications in their work were limited to full rankings of only $n=5$ items, for which the permutation space can be explored exhaustively and, even with a time-consuming inferential process, the MLE is more likely to be achieved. Instead, our implementation of the EM is completely specialized for the Spearman distance, thus exploiting the shortcuts explained in the previous sections.

Concerning publicly available softwares, the \texttt{R} package \texttt{rankdist} \citep{rankdist} performs MLE for the MM in the presence of full and top-$q$ ranking data, including the MMS and its finite mixture extension. However, the analytical simplifications associated to the Spearman distance are not considered, and the MLE solution for $\boldsymbol{\rho}$ is searched locally, slowing down the computational time. In addition, the program generates all the $n!$ permutations to compute the normalizing constant $Z(\theta)$ needed to estimate the concentration parameter, and hence crashes for $n>11$ by limiting the range of MMS applications to few items. In case of top-$q$ rankings the program works with at most $n=7$ items.

The \texttt{pmr} \texttt{R} package \citep{pmr_R} performs frequentist inference on the MMS, but again, it generates all the $n!$ possible permutations, to search the MLE of the consensus ranking globally. As a result, it quickly runs out of memory restricting the MMS applicability to datasets with a maximum of $n=11$ items.

The \texttt{R} package \texttt{BayesMallows} \citep{BayesMallows} implements Bayesian inference for the MMS with any right-invariant distance, including the Spearman, and it also supports partial rankings and mixture models. Even if the procedure scales well with $n$, the implementation of the MMS does not consider the analytical simplifications of the Spearman distance, which could speed up the algorithm considerably. In \texttt{BayesMallows} the normalizing constant of the MM is approximated by means of an importance sampling scheme. In Section \ref{ssec:approx} we compare the performance of our novel approximation of $Z(\theta)$ with different types of approximations, including the one available in \cite{BayesMallows}.

\subsection{Mixture model extension for partial rankings}
\label{ss:EMpartial}

In real applications it could happen that the ranking data are not completely observed. This occurs, 
for instance, when the assessors rank only a subset of the items, typically their top-$q$ preferred ones, or there may be missing data either at random or by design, giving rise to various forms of partial observations. 

In this section we extend the mixture analysis previously described for accommodating inference on partial data. %, following \citep{Busse2007}.
In the spirit of \cite{beckett}, whose goal was to estimate the parameters of the basic MM from partial observations, we generalize its MLE approach based on the EM algorithm by assuming the mixture of MMS as the sampling distribution. We adopt a data completion argument 
%to a full ranking by treating 
that suitably combines the two main sets of latent variables: the missing ranks of the partial sequences and the unobserved group memberships.
%as latent variables, and assuming that the (latent) full rankings are distributed according to a mixture of $\theta-$models. 
%The estimate of the full ranking is
%obtained with the 
%The resulting EM algorithm iterates the estimates of (i) the group membership probabilities, given the MM mixture parameters and the current value of the full ranking
%frequencies; (i) the MM mixture parameters, given the current value of full ranking
%frequencies and group membership probabilities; (ii) the full ranking frequencies conditionally on the current parameter estimates and group membership probabilities. 

%Using the above procedure requires latent variables to account for the missing ranks, in addition to the latent assignment variables required by the mixture model. 
%We therefore expect that the EM will converge slowly, as the convergence speed of EM algorithms depends on the proportion of latent variables \citep{mclachan}. \cm{is this claim necessary for our work?}

Suppose that $N_l$ respondents provide a partial sequence $\boldsymbol r_l$ where only a subset $\mathcal{I}_{l} \subseteq \{1,2,\dots,n\}=\mathcal{I}$ of $n_l=\vert\mathcal{I}_{l}\vert$ items are actually ranked.
%, giving them top ranks from $1$ to $n_j=\vert\mathcal{I}_{j}\vert$. 
As a consequence, the position $r_{li}$ is known only if $i \in \mathcal{I}_{l}$. 
We stress that, in our setting, arbitrary types of censoring patterns are allowed, implying that the set of the assigned positions may be any not-empty subset of the first $n$ integers, leading to  $n_l=1,\dots,n-1$. Moreover, a \textit{full ranking} $\tilde{\boldsymbol{r}}_l$ is defined as the complete sequence obtained by combining the $n_l$ observed entries $\boldsymbol r_l$ with the latent unassigned $n-n_l$ ranks for the items $i' \in \mathcal{I}\setminus\mathcal{I}_{l}$. The latent components of the partially ranked vector are inferred through the EM algorithm, that exploits the information provided by the entire sample to fill the missing entries of the sequence coherently with the observed ones.
%in a way which is compatible with the rest of the data. 

Let $\mathcal{C}(\boldsymbol r_l)\subset \mathcal{P}_n$, be the set of full rankings which are compatible with the partial sequence $\boldsymbol{r}_l$. This means that $\mathcal{C}(\boldsymbol r_l)$ collects all the complete ranked vectors on which the application of a certain censoring pattern yields the partial observation $\boldsymbol{r}_l$. Under the mixture model, the conditional probability of observing the full ranking $\tilde{\boldsymbol{r}}_m\in\mathcal{C}(\boldsymbol{r}_l)$ is

\begin{equation}
\begin{split}
 \label{eq:ProbAug}
p_{r_l}(\tilde{\boldsymbol{r}}_m)&\defeq\mathbb{P}(\tilde{\boldsymbol{r}}_m\,\vert\boldsymbol{r}_l,\boldsymbol{\rho},\boldsymbol\theta,\boldsymbol\omega)\\
&=\frac{\sum\limits_{g=1}^G \omega_g e^{-2\theta_g\left(c_n-{\boldsymbol{\rho}^T_g}\tilde{\boldsymbol{r}}_m\right)-\log Z(\theta_g)}}{\sum\limits_{\tilde{\boldsymbol{s}}\in \mathcal{C}(\boldsymbol{r}_l)}\sum\limits_{g=1}^G \omega_g e^{-2\theta_g\left(c_n-{\boldsymbol{\rho}^T_g}\tilde{\boldsymbol{s}}\right)-\log Z(\theta_g)}}.
\end{split}
 \end{equation}
%Let us denote with 
%$$f(\boldsymbol r)=\sum_{j=1}^N\mathbb{1}_{[r_j=r]}$$
Equality \eqref{eq:ProbAug} extends expression (6) in \cite{beckett} to the finite mixture case. The observed frequency $N_l$ of respondents reporting each partial ranking $\boldsymbol{r}_l$
is then distributed over the full rankings
%in $\mathcal{C}(\boldsymbol r)$ 
that may give rise to
%the partial ranking 
$\boldsymbol{r}_l$, in order to obtain the E-step estimate of the frequency of $\tilde{\boldsymbol{r}}_m$ as
%complete ranking as 
\begin{equation}
\label{eq:eq7beckett}
%    \tilde{f}(\tilde{\boldsymbol r}) = \sum_{\boldsymbol r\,:\,\tilde{\boldsymbol{r}}\in\mathcal{C}(\boldsymbol{r})} f(\boldsymbol r)p_{r}(\tilde{\boldsymbol{r}}).
    \hat{N}_m = \sum_{l:\,\tilde{\boldsymbol{r}}_m\in\mathcal{C}(\boldsymbol{r}_l)} N_l p_{r_l}(\tilde{\boldsymbol{r}}_m).
\end{equation}
The above sum accounts for the fact that, under varying censoring patterns, the same full ranking $\tilde{\boldsymbol r}_m$ can be compatible with different partial rankings; for instance, $\tilde{\boldsymbol r}=(1, 2, 4, 3)$ is compatible with both $\boldsymbol{r}_1=(1, 2, NA, NA)$ and $\boldsymbol{r}_2=(1, NA, NA, NA)$. Obviously, if the observed data share the same censoring pattern (e.g. they are all top-3 implying $n_l=3$ for all $l=1,\dots,L$), the sum in \eqref{eq:eq7beckett} can be avoided and the computation simplifies greatly. 

In practice, the EM scheme detailed in Section \ref{ss:EM} is adapted for handling partial rankings by adding a further E-step, needed to account for the latent ranked entries, and by considering the augmented data $\tilde{\boldsymbol{r}}_m\in\mathcal{C}(\boldsymbol r_1)\cup\cdots\cup\mathcal{C}(\boldsymbol r_L)$ as the sample to draw inference on.

\section{A new approximation of the partition function}
\label{ssec:approx}
%\mc{Marta and Valerio take care of this, also comparing it with other approximations}.

If, on the one hand, the existence of the closed-form for the MLE of $\boldsymbol\rho$ avoids the computationally demanding local search for inferring the consensus ranking, on the other, considering large values of $n$ affects the calculation of $Z(\theta)$ which is needed in the estimation steps of (i) the concentration parameter (see eq. \eqref{e:thetaest}), (ii) the posterior membership probabilities %(needed in the E-step of Section \ref{ss:EM}) 
and (iii)  the conditional probabilities of observing the full rankings compatible with the partial information (see eq. \eqref{eq:ProbAug}). In fact, the rapidly-increasing size $n!$ makes the exhaustive enumeration of the sequences in $\mathcal{P}_n$, and hence the direct calculation of $Z(\theta)=\sum_{\boldsymbol{r} \in \mathcal{P}_{n}} e^{-2\theta\, \left(c_n-\boldsymbol{e}^T\boldsymbol{r}\right)}$,  feasible only for small $n$.
%The number of terms in the sum makes direct calculation $Z(\theta)$ unfeasible for all but very small values of $n$. 
%As a consequence, the MMS did not receive much attention in the past 
%\vale{being preferred to?\cm{
%and has been less explored than the  
%...va meglio?}}being preferred to other 
%MMs with other metrics, in particular with the Kendall, Cayley and Hamming distances. For these metrics, $Z(\theta)$ has a favourable closed-form expression as a simple function depending only on the $\theta$ and $n$ \citep{IrurozkiThesis}.

%In recent years, different approximation strategies have been proposed \citep[see e.g.][]{McCullagh1993, mukherjee2016, vitelli18}, allowing inference even in the case of  a large number $n$ of items. 
In this section, we propose a novel approximation of the distribution of the Spearman distance which enables inference on rankings of a large number $n$ of items.
In fact, this approximation, which may be interesting in its own right (e.g. for sampling procedures), 
%described in Section \ref{ss:appr}, 
is then used to approximate both
%the partition function 
$Z(\theta)$
%of the MMS 
and
%the expected Spearman distance 
$\text{E}_\theta[D]$.

\subsection{The new approximation in practice}\label{ss:appr}
Regardless of the metric adopted in the MM, the partition function 
%$Z(\theta)=\sum_{\boldsymbol{r} \in \mathcal{P}_{n}} e^{-\theta \, \n{\mathbf{1}_n- \boldsymbol{r}}}$, with $\mathbf{1}_n=(1,2,3,\ldots,n)$,
can be more conveniently
%more conveniently 
expressed as 
\begin{equation}\label{eq:zeta}
Z(\theta)=\sum_{d_n\in\mathcal{D}_n}N_{d_n} \,e^{-\theta d_n},
\end{equation}
where $N_{d_n}=\vert\{\boldsymbol{r}\in\mathcal{P}_{n}\,:\,d(\boldsymbol{r},\boldsymbol{e})=d_n\}\vert$ is the number of rankings at distance $d_n$ from 
%the identity permutation 
$\boldsymbol{e}$
%=\left( 1,2,\dots,n \right)$
and $\mathcal{D}_n$ is the set of possible  distance values among the $n$-dimensional rankings.
%\citep{irurozki2016permallows}.
%Thus, for the computation of $Z(\theta)$, one needs to compute 
In the case of Spearman distance,
$\mathcal{D}_n=\{d_n=2h\,: \, h\in\mathbb{N}_0\text{ and } 0\leq d_n\leq d_{\text{max},n}=2\binom{n+1}{3}\}$,
%, that is, the possible Spearman distance values among  $n$-dimensional rankings are all even numbers from 0 to $d_{\max,n}=2\binom{n+1}{3}$, 
implying that the number of terms in \eqref{eq:zeta} is of order $n^3$ rather than $n!$ as in the na\"ive representation of $Z(\theta)$ related to \eqref{eq:ProbMassMallows}. The values $N_{d_n}$
%for all values $d\in\mathcal{D}_n$, which basically 
indicate the frequency distribution of the Spearman distance under the uniform 
%probability over the permutation space 
(null) model and correspond to the sequence A175929 available only up to $n\leq 14$ in the OEIS \citep{sloane17, vitelli18}.
We here propose a strategy to approximate $N_{d_n}$ when $n>14$. 

\begin{figure}[t]
    \centering
    \includegraphics[width=0.5\textwidth]{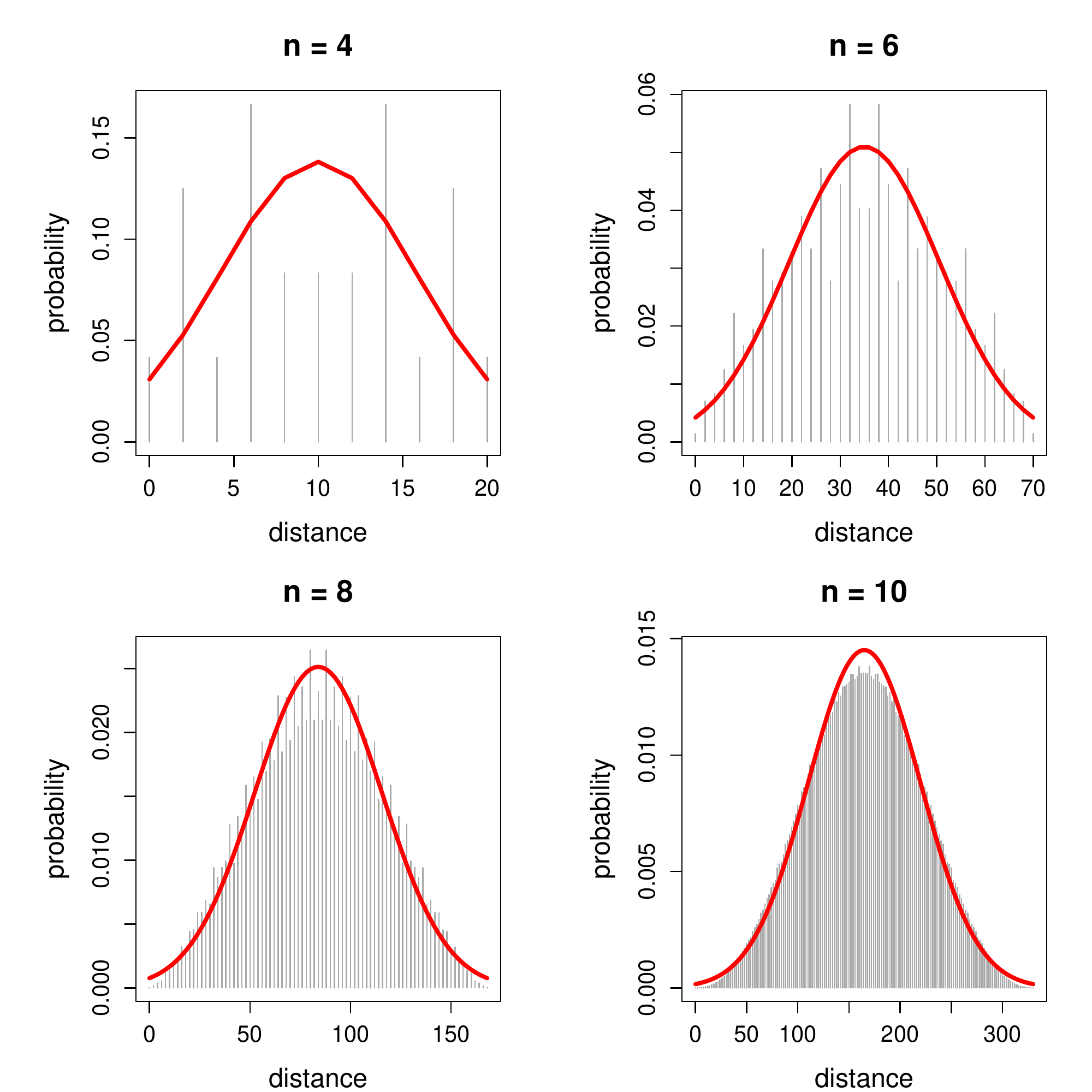}
\caption{Comparison between the probability mass distribution of the Spearman distance under the null model (gray) and the limiting normal density (red) for increasing $n$ values.} 
%fatta con Spear_foot_approx.R
%rifatta con Approssimazione_Card_new.R--> sez3.1
\label{fig:appr}
\end{figure}

As proven in \cite{kendall1970rank}, under the null model the Spearman distance %has mean and variance given by
%$$\text{E}[D_S]=\frac{n^3-n}{6}\qquad\qquad \text{V}[D_S]=\frac{n^2(n-1)(n+1)^2}{36}$$ 
%and, for $n\rightarrow+\infty$, it 
is asymptotically normally distributed (see Figure \ref{fig:appr}).
However, the convergence is slow and not uniform over the domain of the distribution. In particular, even with $n$ large enough, for the distribution to be reliably approximated in the bulk, the tail behaviour will be different from the one predicted by a normal distribution.

To usefully approximate the distribution of extreme values, we could in principle use large deviation theory \citep{cramer1938nouveau,varadhan1966asymptotic} but, as explained in \cite{mukherjee2016}, the computation of explicit results is often not feasible. We can nevertheless exploit general properties of the large deviation rate function, and what we know on the values of $N_{d_n}$ for $n\leq 14$, to obtain an approximation.
As a matter of fact, in \cite{mukherjee2016} it is proved the existence and differentiability of the (scaled) cumulant generating function,\footnote{In the original notation, the author proved the existence and differentiability of the function $Z\left(f, \theta \right) = \lim_{n\to \infty} \frac{1}{n}\log \text{E}\left[ \exp(n\, \theta\, \nu_{\pi}\left[ f\right])\right]$. By setting $f(x,y)=(x - y)^2$ and $\theta=k$, we obtain the desired result.}
$$
\lambda\left(k\right) \defeq \lim_{n\to \infty} \frac{1}{n} \log \text{E}\left[e^{k n x_n}\right] \quad k\in \mathbb{R},
$$
% E_{\mathcal{P}_n}
with $x_n = \frac{d_n}{d_{\max,n}}$. This result allows us to apply the G{\"a}rtner-Ellis theorem \citep{gartner1977large,ellis1984large}, to obtain the approximate rate function
% is finite and differentiable for every $k$, so that we can express the probability of having a (rescaled) distance $x_n  = \frac{d}{n} $ as:
\begin{equation}
    \xi\left( x_n \right) \approx \frac{1}{n} \log\left(\frac{N_{d_n}}{n!} \right),
\end{equation}
with $\xi\left( x_n \right)$ only weakly dependent on $n$ for \mbox{$n\gg 1$.} 
Making use of the known values of $N_{d_n}$ for $n\leq 14$, we can derive for $\xi\left( x_n \right)$ the form
\begin{equation}
\label{approssimazione}
    \xi\left(x_n \right) = a_0 +  a_1 \log\left[ x_n \left( 1- x_n \right)\right] +  a_2  x_n \left( 1- x_n \right).
\end{equation}
The formal derivation of \eqref{approssimazione} and the value of the coefficients $(a_0, a_1, a_2)$ are detailed in Supplementary Material A.
With this result, our approximation of $N_{d_n}$ for $8<d_n<d_{\max,n}-8$ is \begin{equation}\label{eq:freq}\hat{N}_{d_n}\propto n!\exp\left[n\,\xi\left(x_n \right)\right].\end{equation} 
The above approximation is used for $n>14$.
The values of ${N}_{d_n}$ for $d_n=0,2,4,6$ (as well as those for their symmetrical $d_n=d_{\max,n},d_{\max,n}-2,d_{\max,n}-4$ and $d_{\max,n}-6$) follow from combinatorial facts and are equal respectively to $1,(n-1),\binom{n-2}{2} $ and $(n-2)^3/6-(n-2)^2+ 23(n-2)/6-1$.

%their  the first four cases of \eqref{eq:freq} follow from combinatorial facts.

%\begin{equation}\label{eq:freq}\footnotesize
%\hat{N}_{d}=\begin{cases}
  %    1, & \text{if}\ d=0,d_{\max,n} \\
   %   n-1, & \text{if}\ d=2, d_{\max,n}-2\\
    %  \binom{n-2}{2}, & \text{if}\ d=4, d_{\max,n}-4\\
     % (n-2)^3/6-(n-2)^2+ 23(n-2)/6-1, & \text{if}\ d=6, d_{\max,n}-6\\
      %\frac{(n-2)}{6}[(n-2)^2-6(n-2)+23]-1, & \text{if}\ d=6, d_{\max,n}-6\\
      %n!\exp\left[n\,\xi\left(x_n \right)\right]%\left(a_0 + a_1\log\vertx_n(1-x_n)\vert + a_2\vertx_n(1-x_n)\vert\right) 
     % & \text{otherwise.}
    %\end{cases}
%\end{equation}
%\hspace{0.5cm}
 
\begin{figure}[t]
    \centering
    \includegraphics[width=0.5\textwidth]{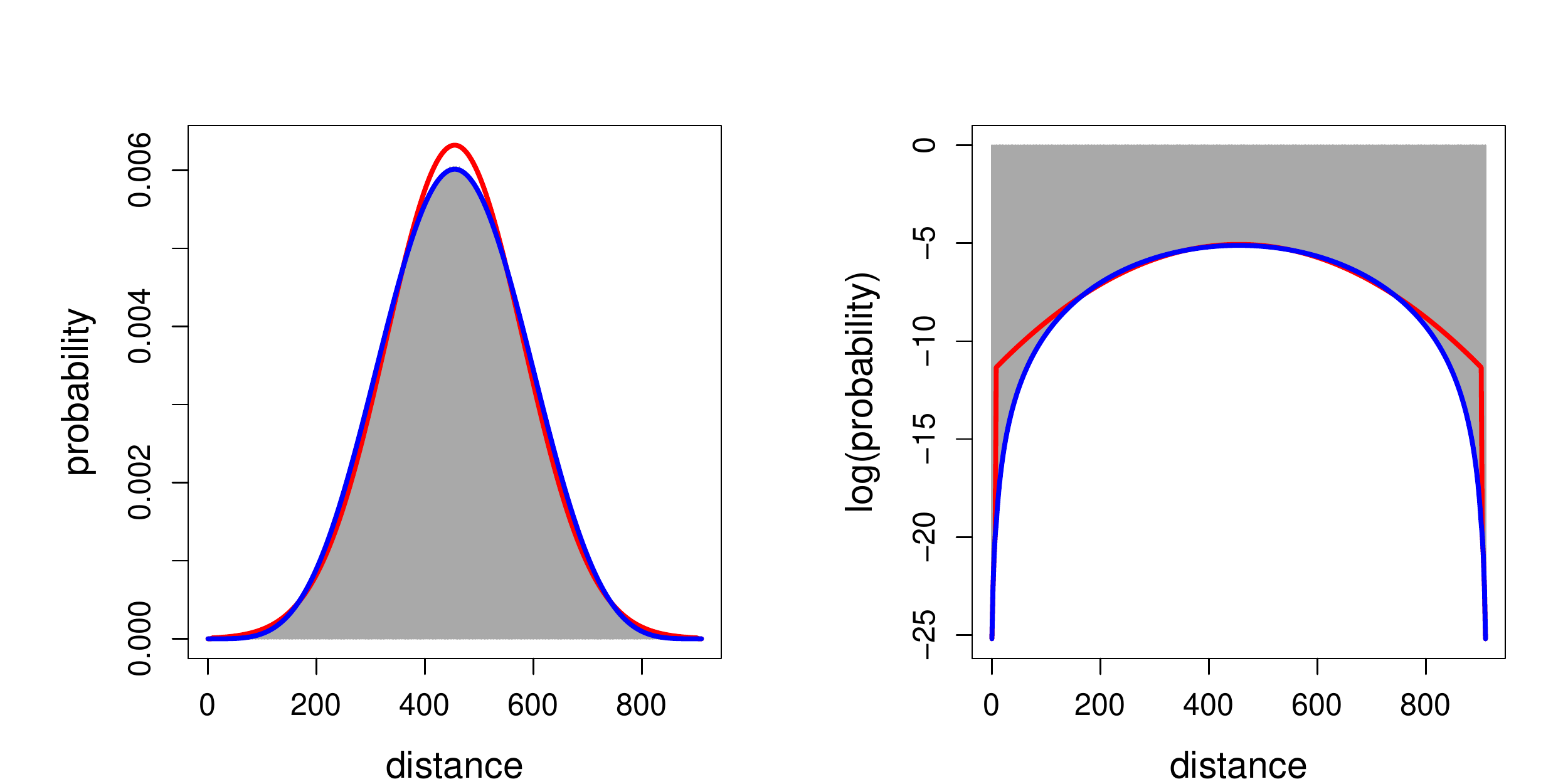}
\caption{Comparison between the limiting normal density (red line) of the probability distribution of the Spearman distance under the null model (gray), and the density obtained with our new approximation (blue line). Left: original scale. Right: logarithmic scale. $n=14$.} %fatta con Approssimazione_Card_new.R --> sez3.1
\label{fig:appr2}
\end{figure}
Figure \ref{fig:appr2}  displays the  probability mass distribution of the Spearman distance for $n=14$ (in gray), its limiting normal density (red line) and the density obtained with our new approximation (blue line) in the original scale (left panel) and in the logarithmic scale (right panel) for better visualization. 

Equipped with the approximation \eqref{eq:freq}, we can 
%immediately obtain an 
approximate the partition function \eqref{eq:zeta} and $\text{E}_{\theta}[D] =\sum_{d_n\in\mathcal{D}_n}dN_{d_n} \,e^{-\theta d_n}/\sum_{d_n\in\mathcal{D}_n}N_{d_n} \,e^{-\theta d_n}$
%$\hat Z^\prime(\theta)$ 
%needed, respectively, in (i) the E-step, for both estimating the membership probabilities and the full ranking frequencies, and (ii) the M-step, for estimating the concentration parameter $\theta$. These approximations can be obtained 
by simply replacing the exact $N_{d_n}$ with $\hat{N}_{d_n}$.
%in \eqref{eq:zeta} to address point (i), and in \eqref{e:thetaest} to address point (ii), by noting that
%$$\text{E}_{\theta}[D_S] = \frac{\sum_{d\in\mathcal{D}_n}dN_{d} \,e^{-\theta d}}{\sum_{d\in\mathcal{D}_n}N_{d} \,e^{-\theta d}}.$$

%$$\text{E}_{\theta_g}[D_S] = \frac{\sum_{d\in\mathcal{D}_n}dN_{d} \,e^{-\theta_g d}}{\sum_{d\in\mathcal{D}_n}N_{d} \,e^{-\theta_g d}}.$$

\begin{figure*}[h!]
    \centering
    \includegraphics[width=\textwidth]{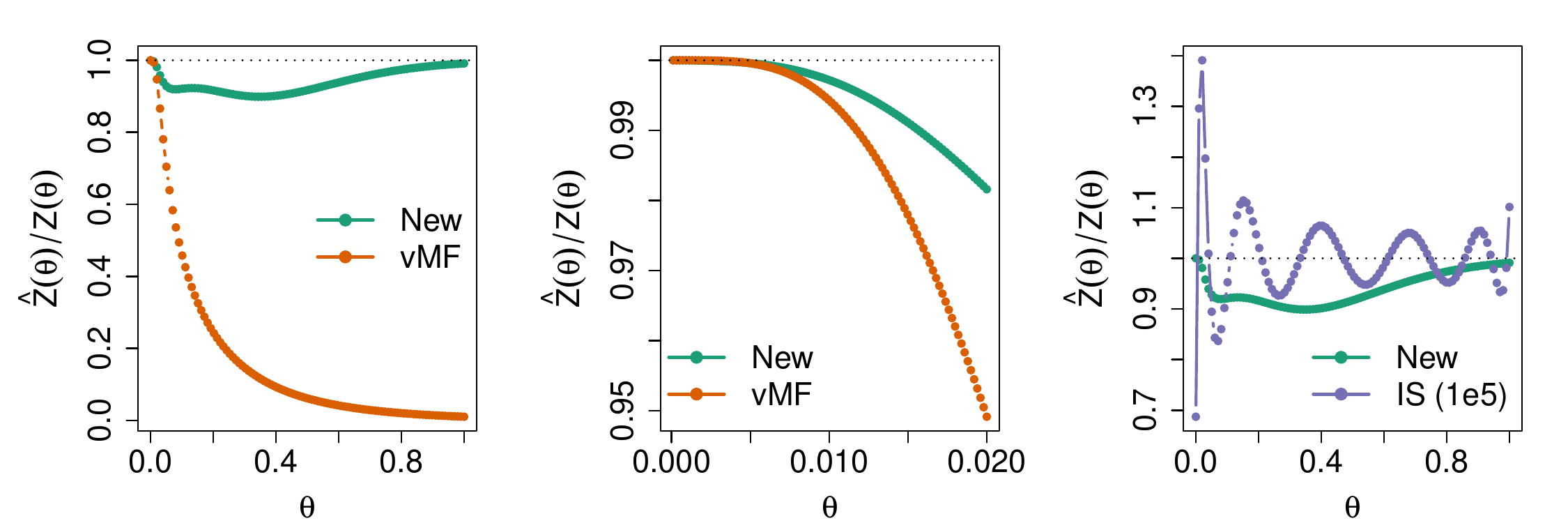}%fatta con fig_approx.R-->sez3.2
\caption{Ratio between alternative approximations of the partition function and the exact one for $n=14$ and $\theta\in(0,1]$ (left and right) and $\theta\in(0,0.02]$ (middle).} 
%\caption{Ratio of the approximate partition function computed via different methods to the exact, $\hat{Z}(\theta)/Z(\theta)$, as a function of $\theta$, with $n=14$. Left and middle panel: $\hat{Z}_\text{New}/Z(\theta)$, and $\hat{Z}_\text{vMF}/Z(\theta)$ against the interval $\theta\in(0,1]$ and $\theta\in(0,0.02]$ respectively; right panel: $\hat{Z}_\text{New}/Z(\theta)$, and $\hat{Z}_\text{IS}/Z(\theta)$ against the interval $\theta\in(0,1]$.} 
\label{tab:compZ}
\end{figure*}

\subsection{A comparison with other solutions}\label{sec:compZ}

In this section we compare our new approximation $\hat Z_\text{New}(\theta) = \sum_{d_n\in\mathcal{D}_n}\hat{N}_{d_n} \,e^{-\theta d_n}$ with two others % approximations of the partition function 
proposed in the literature, one analytical and the other simulation-based. The former, based on the von Mises-Fisher (vMF) distribution and originally proposed in \cite{McCullagh1993}, has been recently exploited by \cite{XuAlvoYu2018} for inference in their angle-based model for rankings and is given by
%, which is a transformation of the MMS. 
%In \citep{McCullagh1993} is noted that the MMS is essentially equivalent to the von Mises-Fisher model on the sphere (except for the discrete nature of the sample space). Therefore,  the normalization constant from the von Mises-Fisher distribution can be used as approximation for the partition function of the MMS. 
%The closed-form of the approximation is given by
$$Z(\theta)\approx \hat{Z}_\text{vMF}(\theta)=
\frac{2^\frac{n-3}{2} n! I_\frac{n-3}{2}(\kappa)\Gamma(\frac{n-1}{2})}{\kappa^\frac{n-3}{2}e^{\kappa}},$$
where $\kappa={\theta n(n^2-1)}/6$, $I_x(\cdot)$ is the modified Bessel function of the first kind with order $x$, and $\Gamma(\cdot)$ is the Gamma function. The reader is referred to Supplementary Material B for the formal derivation of $\hat{Z}_\text{vMF}(\theta)$ in the context of (non-standardized) rankings.
The latter $\hat Z_\text{IS}(\theta)$, based on the Importance Sampling (IS) technique, is due to \cite{vitelli18} to which we refer for details. 

In Figure \ref{tab:compZ} we plot the ratio $\hat{Z}(\theta)/Z(\theta)$ as a function of $\theta$ for $n=14$. In the left and middle panels, where the approximations $\hat{Z}_\text{New}$ (in green) and $\hat{Z}_\text{vMF}$ (in orange) are compared over the interval $\theta\in(0,1]$ and $\theta\in(0,0.02]$ respectively, we notice that for small values of $\theta$, the vMF approximation and our proposal perform both very good (middle panel). However, the vMF approximation  deteriorates fast already for $\theta>0.015$ and its negative bias rapidly becomes prominent (left panel). This happens because the vMF approximation is an integral over the entire ($n-1$)-dimensional sphere enclosing all the points of $\mathcal{P}_n$ and, for growing $\theta$, the permutations are more and more concentrated in a small region of this sphere under the MMS, implying that $\hat{Z}_\text{vMF}$ is strongly biased. On the contrary, our new approximation performs very well for increasing values of $\theta$, because it exploits the knowledge of the true values of $N_{d_n}$ available for small $d_n$, which are the ones that mostly matter in  \eqref{eq:zeta}.

Instead, the IS approximation has an oscillating behavior around the exact, and its error is generally small even if, when compared to our new approximation, it is closer to the exact for few values of $\theta$. Note that, even if the performance of the IS is generally good, being based on random sampling, it is  computationally very demanding, while our new approximation is immediately available for any value of $n$. %we expect that the IS approximation works well for small values of $\theta$, when exploring the sample space is easier % \footnote{Sampling is difficult when a distribution is highly concentrated, because many iterations are needed to sample rare observations.} 
%and, for the same reason, it will need many more iterations to have the same error rate when $n$ is larger.

From the above results we evince that our proposed new approximation performs very good for a wide range of values of $\theta$, not only for small ones, like the analytical approximation resorting on the vMF distribution. This is particularly relevant when dealing with mixture models, where the cluster-specific concentrations %parameters within the mixture components 
typically grow with the number of clusters detected.

To complete the study of the proposed approximation, in Supplementary Material C we document that its effect on the inference on  $\theta$  is generally negligible.

\section{Simulation studies}\label{sec:simu}
%In this section, we present the inferential results of our EM algorithm on tree types of simulated data: i) full and homogeneous rankings; ii) full and heterogeneous rankings and iii) partial and homogeneous rankings.

\subsection{Experiments with different sample sizes and number of items}\label{ssec:simu1}
%Compare with \texttt{rankdist} \citep{rankdist} until it does not run out of memory?. \mc{non capisco se loro stimano o no $\theta$, non riesco a trovarlo nel loro output}.

The first simulation study considers data drawn from the homogeneous MMS ($G^*=1$) under different parameters settings, and then evaluates the performance of our fitting procedure in terms of model parameters estimates with multiple criteria. Specifically, for each combination of $(N,n)$ with sample size $N\in\{50, 200, 500, 1000\}$ and number of items $n\in\{5,10,14,15,25,50,100\}$, we  independently simulate 100 datasets from the homogeneous MMS, by randomly choosing the true consensus ranking as $\boldsymbol\rho\sim\text{Unif}\{\mathcal{P}_n\}$ and sampling uniformly $\theta\sim\text{Unif}(\theta_\text{min},\theta_\text{max})$ over different intervals depending on the value of $n$ (see Table \ref{tab:theta_simu}, which collects the bounds of the uniform sampling intervals). In this way, we have a comparable concentration over the permutation set in the samples with different values of $n$. %MMS defined over a permutation set $\mathcal{P}_n$ with a varying dimension. % and the Appendix for further discussions on the rationale behind these choices). 
%We consider two different scenarios: in scenario 1 the concentration parameter is sampled uniformly in  $\theta\sim\text{Unif}(\theta_\text{min}=0.1/n,\theta_\text{max}=5/n)$; in scenario 2 the concentration parameter is sampled uniformly in narrower intervals corresponding to smaller values of $\theta$. Table \ref{tab:theta_simu} collects  the extreme points of the intervals for sampling $\theta$.\footnote{\mc{The ratio behind these choices is explained in the Online Appendix.}}

%As a consequence, the reconstruction of the model parameters is more difficult under scenario 2 than under scenario 1. 

\begin{table}[h!]
\centering
\begin{tabular}{lrrrHrrr}
%  & \multicolumn{7}{c }{Number of items}  \\
% \cline{2-8}
 \hline
$n$ & 5 & 10 & 14,15 & 15 & 25 & 50 & 100 \\
  \hline
  \hline
$\theta_\text{min}$ & .15 & .05 & .025 & .025 & .010 & .002 & .0002 \\
$\theta_\text{max}$ & .30 & .10 & .050 & .050 & .020 & .010 & .0020 \\
  \hline
\end{tabular}

%}
\caption{Bounds of the intervals for the uniform sampling of $\theta$ for different number of items. % under the two scenarios considered.
}\label{tab:theta_simu}
\end{table}

%\begin{table}[h!]
%\centering
%\footnotesize
%% \subcaptionbox{Scenario 1}{
%% \begin{tabular}{lrr}
%%   \hline
%%  $n$ & $\theta_\text{min}$ & $\theta_\text{max}$ \\
%%  \hline
%% 5 & 0.02 & 1 \\ 
%%   10 & 0.01 & 0.5 \\ 
%%   14 & 0.0071 & 0.36 \\ 
%%   15 & 0.0067 & 0.33 \\ 
%%   25 & 0.004 & 0.2 \\ 
%%   50 & 0.002 & 0.1 \\ 
%%   100 & 0.001 & 0.05 \\ 
%%   \hline
%% \end{tabular}
%% }
%%\hspace{0.5cm}
%%\subcaptionbox{Scenario 2}{
%\begin{tabular}{lrr}
%  \hline
% $n$ & $\theta_\text{min}$ & $\theta_\text{max}$ \\ 
%  \hline\hline
%5 & 0.15 & 0.3 \\ 
% 10 & 0.05 & 0.1 \\ 
% 14 & 0.025 & 0.05 \\ 
% 15 & 0.025 & 0.05 \\ 
% 25 & 0.01 & 0.02 \\ 
% 50 & 0.002 & 0.01 \\ 
% 100 & 0.0002 & 0.002 \\ 
%   \hline
%\end{tabular}

%}
%\caption{Extreme points of the intervals for uniform sampling of $\theta$. % under the two scenarios considered.
%}\label{tab:theta_simu}
%\end{table}

The inferential ability of the proposed EM algorithm to recover the actual parameters values has been evaluated by averaging over the 100 
%synthetic 
datasets the following measures:
\begin{itemize}
    \item %$m_\theta=1-\frac{2}{\hat\theta/\theta+1}$ or $\log(\hat\theta/\theta)$ or 
    $m_\theta=\vert\hat\theta-\theta\vert/\theta$, the relative error between the true and estimated concentration;
    %\item $\phi_{\boldsymbol\theta} = \mathbb{1}(\hat\theta>\theta)$
    \item $m_{\boldsymbol\rho}=d(\hat{\boldsymbol\rho},\boldsymbol\rho)/d_{\max,n}$, 
    %where $d_{\max,n} =2\binom{n+1}{3}$, 
    the relative distance between the true and estimated consensus;
    \item $\phi_{\boldsymbol\rho}=\mathbb{1}_{(\boldsymbol\rho=\hat{\boldsymbol \rho})}$, the matching between the true and the estimated consensus;
    \item $\phi_G=\mathbb{1}_{(\hat G= {G^*})}$, the matching between the true and  estimated number of components. The optimal number $\hat G$ of groups is automatically selected by means of the Bayesian Information Criterion, BIC \citep{Schwarz}. In particular, we select $\hat G$ by applying the elbow rule to the screeplot of the BIC values \citep{Zhao08,cite-key}, in place of the more popular choice of selecting the minimum BIC, because of the tendency of the MMS to over-fit the data \citep{MurphyMartin2003, rankdist}.\footnote{Note that this criterion is useful only for automatizing the simulations: when choosing the number of clusters in applications the researcher typically uses the BIC, as well as other relevant measures, as a guidance, and decides the number of groups based also on other heuristic criteria (such as the size of the clusters, their separation, or prior information depending on the application at hand).
    } %\mc{spiegare il criterio che usiamo e perchè. questione dell'overfitting e della tendenza a creare singletons. Supportare anche con MurphyMartin, rankdist e con altri lavori su misture e Mallows. Descrivere la procedura: usiamo il BIC pensalizzato per singletons e la elbow rule (dare almeno una referenza della elbow rule col BIC).  }
\end{itemize}

\begin{table*}[h!]
\centering
\subcaptionbox{$n\leq14$}{
\begin{tabular}{Hllrrrr}
  \hline
  &$n$ & $N$ & $m_\theta$ & $m_{\boldsymbol\rho}$ & $\phi_{\boldsymbol\rho}$ & $\phi_G$ \\   \hline\hline
  1 & 5 & 50 & .078 & .0000 & 1.00 & 1.00 \\ 
  2 & 5 & 200 & .045 & .0000 & 1.00 & 1.00 \\ 
  3 & 5 & 500 & .034 & .0000 & 1.00 & 1.00 \\ 
  4 & 5 & 1000 & .020 & .0000 & 1.00 & 1.00 \\ \hline
  5 & 10 & 50 & .071 & .0039 & .54 & .87 \\ 
  6 & 10 & 200 & .034 & .0002 & .97 & .91 \\ 
  7 & 10 & 500 & .024 & .0000 & 1.00 & 1.00 \\ 
  8 & 10 & 1000 & .016 & .0000 & 1.00 & 1.00 \\ \hline
  9 & 14 & 50 & .061 & .0063 & .08 & .99 \\ 
  10 & 14 & 200 & .031 & .0010 & .63 & 1.00 \\ 
  11 & 14 & 500 & .019 & .0001 & .98 & .98 \\ 
  12 & 14 & 1000 & .015 & .0000 & 1.00 & 1.00 \\ \hline
  \end{tabular}
 }
\hspace{.5cm}
\subcaptionbox{$n>14$}{
\begin{tabular}{Hllrrrr}
  \hline
  $n$ & $N$ & $m_\theta$ &  $m_{\boldsymbol\rho}$ & $\phi_{\boldsymbol\rho}$ & $\phi_G$ \\   \hline\hline
  13 & 15 & 50 & .062 & .0050 & .06 & 1.00 \\ 
  14 & 15 & 200 & .030 & .0006 & .72 & 1.00 \\ 
  15 & 15 & 500 & .020 & .0000 & .98 & .99 \\ 
  16 & 15 & 1000 & .015 & .0000 & 1.00 & 1.00 \\ \hline
  17 & 25 & 50 & .040 & .0052 & .00 & 1.00 \\ 
  18 & 25 & 200 & .029 & .0012 & .10 & 1.00 \\ 
  19 & 25 & 500 & .022 & .0002 & .58 & 1.00 \\ 
  20 & 25 & 1000 & .021 & .0000 & .93 & 1.00 \\ \hline
  21 & 50 & 50 & .053 & .0051 & .00 & 1.00 \\ 
  22 & 50 & 200 & .042 & .0013 & .00 & .99 \\ 
  23 & 50 & 500 & .041 & .0005 & .00 & 1.00 \\ 
  24 & 50 & 1000 & .047 & .0002 & .05 & 1.00 \\
  25&50 & 10000 & .042 &  .0000 & .98 & 1.00 \\ \hline
 26&100 & 50 & .104 & .0126 & .00 & 1.00 \\ 
 25&100 & 200 & .083 &  .0026 & .00 & 1.00 \\ 
 25&100 & 500 & .091 &  .0018 & .00 & .93 \\ 
 25&100 & 1000 & .093 & .0006 & .00 &1.00 \\  \hline
 \end{tabular}}
\caption{Inferential performance of the EM algorithm on simulated full rankings.
}\label{tab:fit2}
\end{table*}

In Table %s \ref{tab:fit1} and 
\ref{tab:fit2} we report the results of the fit. %for the two scenarios considered.
For all $n$, we see that 
%the ability to correctly estimate the modal consensus ranking $\boldsymbol\rho$ consistently 
all performance criteria improve with $N$
%($m_{\boldsymbol\rho}$ decreases and $\phi_{\boldsymbol\rho}$ increases), 
and are generally very well, at least for $n\leq 25$. As a matter of fact, the performance of the fitting procedure is linked to the sparsity of the problem: when the data are too sparse (that is, with large $n$ and small $N$%\footnote{Note that $\boldsymbol\rho$ is an element of the permutation space, whose dimension grows like $n!$, result that makes the estimation problem critical when $n$ is large and $N$ is not large enough.}
), and the concentration parameter is small, recovering the exact true consensus, measured by $\phi_{\boldsymbol\rho}$, 
becomes very difficult (see the results for $n=50,100$ in Table \ref{tab:fit2}(b)). However, $m_{\boldsymbol\rho}$ is very small even in these extreme cases, testifying a general good performance of the procedure. In addition, it is enough to sufficiently enlarge the sample size to recover exactly the true $\boldsymbol\rho$ in almost all samples considered (see Table \ref{tab:fit2}(b) for $n=50$ and $N=10000$). Finally, from Table \ref{tab:fit2}, we note also that the inferential ability does not seem to be worsened by the introduction of the approximation of $Z(\theta)$. In fact, $m_\theta$, is comparable for the cases $n=14,15$, the first estimated with the exact $Z(\theta)$, and the second with $\hat{Z}_\text{New}(\theta)$.
%(see also Sections \ref{ss:appr} and \ref{sec:compZ}).

%As a final remark, we note that the first criterion for the automatic choice of $G$ (measured by $m_\text{bic,1}$) suffers in particular when the data are sparse, but the automatic selection based on the elbow rule (measured by $m_\text{bic,2}$) performs very well. Note however, that these two criteria are useful only for automatizing the fitting procedure: in true applications the researcher would inspect the BIC plot visually, and decide the number of groups based also on other heuristic criteria (such as the size of the clusters, their closeness, or prior information depending on the application at hand). 

%\mc{questo succede perchè mle non è uguale a true rho!}

%We also report the computational time for each scenario considered.
%We note that the computational time does not increase  with $N$, but only with $n$. The reason for this result lies in the computational simplifications of our procedure which, when no groups are considered, does not need to locally search $\boldsymbol \rho$ in \mathcal{P}_n. 

\subsection{Experiments with heterogeneous data}\label{ssec:simu2}
%simulations_marta_clusters.R

In this section we study the performance of our procedure in terms of its ability to correctly recover the cluster structure.

We simulate 100 datasets from different mixtures of MMS with varying number of clusters $G^*=2,3,4$, while keeping fixed the number of items $n=25$, and the sample size $N=2000$. For each dataset, we sample (i) the mixture weights from a symmetric Dirichlet distribution with large shape parameter to favour populated and balanced clusters, that is, $(\omega_1,\dots,\omega_{G^*})\sim\text{Dirichlet}(\alpha=(2G^*,\dots,2G^*))$; (ii) the cluster consensus rankings $\boldsymbol\rho_1,\dots,\boldsymbol\rho_{G^*}$ in $\mathcal{P}_n$ in order to be at least at Spearman distance $d_{\max,n}/(n-1)=(n^2-1)/3$ from each other;
%\footnote{This corresponds to $d_{\max,n}/(n-1)$.}
 and (iii) the concentration parameters uniformly in different intervals, corresponding to three different scenarios concerning the separation of the clusters from each other (see Table \ref{tab:theta_simu2}). %\mc{In particular, $\theta\sim\text{Unif}(0.002,0.004)$ in the most difficult scenario, $\theta\sim\text{Unif}(0.003,0.005)$ in the medium difficulty one, and $\theta\sim\text{Unif}(0.004,0.006)$ in the low difficulty scenario.} 
%\mc{aggiungere qualcosa sul paper di Liu 2018 forse per giustificare queste scelte.}

\begin{table}[h!]
\centering
\begin{tabular}{lccc}
 & \multicolumn{3}{c}{Cluster separation}\\
 \cline{2-4}
  & {High} & {Medium} & {Low} \\  \hline
  \hline
$\theta_\text{min}$ & .004 & .003 & .002 \\
$\theta_\text{max}$ & .006 & .005 & .004  \\
  \hline
\end{tabular}
\caption{Bounds of the intervals for the uniform sampling of $\theta$ in the three group separation scenarios. % under the two scenarios considered.
}\label{tab:theta_simu2}
\end{table}

%\mc{(see the Appendix for further details on the specific choices for the dispersion parameters). Maybe also report plots which help to visualize the difficulty of the chosen scenarios.}
For each sample, we then fit the data with varying number of components $G = 1,\dots,6$ and select the optimal $\hat G$ by means of the BIC like in the previous section. % (criterion denoted by $m_\text{bic,2}$). 

\begin{table}[h!]
\centering
\begin{tabular}{lrrrrrrrrrr}
%\hline
 & \multicolumn{8}{c }{Cluster separation}\\
 \cline{2-9}
  & \multicolumn{2}{c }{High} & & 
\multicolumn{2}{c }{Medium} & &
\multicolumn{2}{c }{Low} \\
 % \cline{2-3} \cline{5-6} \cline{8-9}
 \hline
 $G^*$ & $\phi_G$ & $\phi_{\boldsymbol{z}}$ &&
 $\phi_G$ & $\phi_{\boldsymbol{z}}$ && 
  $\phi_G$ & $\phi_{\boldsymbol{z}}$\\ 
\hline\hline
 2 & 1                & .00021   && 1.00 & .015 && .99             & .131\\
 3 & 1                & .00016   && 1.00 & .016 && 1.00             & .135\\
 4 & 1             & .00010   && .98  & .012 && .96         & .104\\
\hline
\end{tabular}
\caption{Inferential performance of the EM algorithm on rankings simulated from alternative MMS-mixtures.}\label{tab:bic}
\end{table}

%\begin{table}[h!]
%\centering
%\footnotesize
%\begin{tabular}{cccc}
%\hline
%$\theta$ & $G^*$ & $\phi_G$ & $\phi_{\boldsymbol{z}}$ \\ \hline\hline
%\multirow{3}{*}{Easy} & 2 & 1                & 0     \\ \cline{2-4} 
% & 3 & 1                & 0     \\ \cline{2-4} 
% & 4 & 0.95             & 0     \\ \hline\hline
%\multirow{3}{*}{Medium} & 2 & 1                & 0.015 \\ \cline{2-4} 
%& 3 & 1                & 0.016 \\ \cline{2-4} 
%& 4 & 0.98             & 0.012 \\ \hline\hline
%\multirow{3}{*}{Difficult} & 2 & 0.99             & 0.131 \\ \cline{2-4} 
%& 3 & 1.00             & 0.135 \\ \cline{2-4} 
% & 4 & 0.96             & 0.104 \\ \hline
%\end{tabular}
%\caption{Inferential performance of the EM algorithm on rankings simulated from alternative MMS-mixtures.}\label{tab:bic}
%\end{table}

In Table \ref{tab:bic} we report $\phi_G=\mathbb{1}_{(\hat{G}=G^*)}$ as before, plus the mis-classification rate, defined as the fraction of the sampled units classified in the wrong mixing group, $\phi_{\boldsymbol{z}} = \sum_{j=1}^N\mathbb{1}(z_j\not=\dot{z}_j)/N$, with the group allocation $\dot{z}_j$ determined by the \textit{maximum a posteriori} (MAP) of the posterior membership probabilities. As expected, the ability to correctly reconstruct the clusters decreases as the difficulty of the problem increases. In general, however, we are happy with the results especially because, in the most difficult scenario, the mis-classification rate is always below 15\%. 

\begin{table*}[h!]
\centering
\begin{tabular}{lrHrrHrcrHrrHr}
%\cline{2-14}
& \multicolumn{6}{c }{$n=10$} & & \multicolumn{6}{c }{$n=15$} \\
 \cline{2-7} \cline{9-14}
  Censoring & $m_\theta$ & $\phi_{\boldsymbol\theta}$ & $m_{\boldsymbol\rho}$ & $\phi_{\boldsymbol\rho}$ & $m_\text{bic,1}$ & $\phi_G$ & &
   $m_\theta$ & $\phi_{\boldsymbol\theta}$ & $m_{\boldsymbol\rho}$ & $\phi_{\boldsymbol\rho}$ & $m_\text{bic,1}$ & $\phi_G$ \\   
 
 \hline\hline
full & .0175 & .57 & .0000  & 1.00 & .99 & .99 && .0175 & .33 & .0000  & 1.00 & 1.00 & 1.00 \\
A & .0192 & .53 & .0000  & 1.00 & .25 & .90 && .0176 & .33 & .0000  & 1.00 & .49 & .94 \\  
B & .0199 & .56 & .0002  & .96 & .37 & .83 && .0182 & .21 & .0002  & .90 & .75 & .82 \\ 
\hline
\end{tabular}
\caption{Inferential performance of the EM algorithm on simulated partial rankings.}
 \label{t:SIMpart}
\end{table*}

\subsection{Simulations for incomplete rankings}\label{ssec:simu3}

Through an additional simulation study, we explore the impact of censoring on the inference, by considering
different patterns of missingness of partial rankings drawn from the homogeneous MMS.
To this aim, we first simulate 100 datasets of $N=500$ full rankings for the cases $n=10$ and $n=15$. Specifically, the observations are drawn from the MMS with  $\boldsymbol\rho\sim\text{Unif}\{\mathcal P_n\}$ and $\theta\sim\text{Unif}(0.1,0.15)$ for $n=10$ and $\theta\sim\text{Unif}(0.5,0.6)$ for $n=15$. Then, partial top-rankings are obtained by randomly censoring all the complete sequences according to two patterns of truncation, labelled respectively as A and B and characterized by an increasing rate of missingness of the items ranked in the bottom positions. In both censoring scenarios, the random number of censored bottom positions varied in $\{2,3,4,5,6\}$, but it was sampled  by using probability masses $\{0.1,0.1,0.1,0.1,0.6\}$ under the censoring scenario A and $\{0.1,0.2,0.4,0.2,0.1\}$ under censoring scenario B.

% Tue Dec 28 16:24:20 2021
%\begin{table}[h!]
%\centering
%\footnotesize
%\begin{tabular}{rrrHrrHr}
 % \hline
 %$n$ & censoring & $m_\theta$ & $\phi_{\boldsymbol\theta}$ & $m_\rho$ & $\phi_{\boldsymbol\rho}$ & $m_\text{bic,1}$ & $m_\text{bic,2}$ \\   \hline\hline
%10 & full & 0.0175 & 0.57 & 0.0000  & 100 & 0.99 & 0.99 \\ 
%10 & A & 0.0192 & 0.53 & 0.0000  & 100 & 0.25 & 0.90 \\ 
%10 & B & 0.0199 & 0.56 & 0.0002  & 96 & 0.37 & 0.83 \\ 
 %  \hline
%15 & full & 0.0175 & 0.33 & 0.0000  & 100 & 1.00 & 1.00 \\
%15 & A & 0.0176 & 0.33 & 0.0000  & 100 & 0.49 & 0.94 \\ 
%%15 & B & 0.0218 & 0.21 & 0.0005  & 70 & 0.68 & 0.64 \\ 
%15 & B & 0.0182 & 0.21 & 0.0002  & 90 & 0.75 & 0.82 \\ 
%\hline
%\end{tabular}
%\caption{Simulation study on partial top-rankings.}
% \label{t:SIMpart}
%\end{table}

From Table \ref{t:SIMpart} 
%shows the inferential performance measures resulting from the analysis of the full rankings and the censored ones. As expected, 
we note that the ability to recover the exact parameter value, as well as the proportion of times that the homogeneous MMS is correctly selected, %in favor of the heterogeneity assumption, 
decreases as the rate of missingness increases. However, for both censoring scenarios, all the criteria exhibit very good values when compared to those obtained in the full ranking analysis, testifying the effectiveness of our EM procedure to draw inference from partial observations.

% \begin{table*}[h!]
% \centering
% \begin{tabular}{cccccc}
% \hline
%  $G=1$ & $G=2$ & $G=3$ & $G=4$ & $G=5$ & $G=6$ \\
% \hline
% \hline
% 16437.9 & 15873.8 & \textbf{15814.1} & 15803.0 & 15768.6 & 15747.3  \\

% \hline
% \end{tabular}
% \caption{BIC values of the MMS fitted to the reading genres data.}
%  \label{t:BICreading}
% \end{table*}
% % 

\section{Applications}\label{sec:applications}

\subsection{Analysis of the sports data}

We revisit the well-known sports  benchmark data \citep{Marden1995}, available in the \texttt{R} package \texttt{Rankcluster} \citep{Rankcluster}, which collects $N=130$ full rankings. These represent the preferences of students at the University of Illinois towards $n=7$ sports, namely Baseball, Football, Basketball, Tennis, Cycling, Swimming and Jogging.
We explored the presence of a cluster structure in the sample through the estimation and comparison of alternative mixture models with $G=1,\dots,5$,
%and by considering different ranking distributions to model the groups within each mixture, 
and components given by: 1) the MM with Spearman distance (MMS); 2) the MM with Kendall distance (MMK); 3) the MM with Cayley distance (MMC); 4) the MM with Hamming distance (MMH) and 5) the Plackett-Luce model (PL).
\begin{table}[h!]
\centering
\begin{tabular}{lccccc}
%\hline
\hline
& MMS & MMK & MMC & MMH & PL  \\
\hline
\hline
$\hat{G}$ & $2$ & $2$ & $2$ & $1$ & $2$  \\
%\bf{2144.54} & 2156.40 & 2204.32 & 2206.43 & 2195.47\\
%BIC & \bf{2144.54} & 2155.09 & 2190.22 & 2206.43 & 2195.47\\
BIC & \bf{2144.5} & 2155.1 & 2190.2 & 2206.4 & 2195.5\\
\hline
\end{tabular}
\caption{BIC and number of groups for the alternative mixtures fitted to the sports data.}
 \label{t:BICsports}
\end{table}
%

%
%\begin{itemize}
%\item[-] the MM with the Spearman distance (MMS-mix);
%\item[-] the MM with the Kendall distance (MMK-mix);
%\item[-] the MM with the Cayley distance (MMC-mix);
%\item[-] the MM with the Hamming distance (MMH-mix);
%\item[-] the Plackett-Luce model (PL-mix).
%\end{itemize}
%
Table \ref{t:BICsports} shows the best fitting mixture under each parametric scenario, and the corresponding estimated number $\hat{G}$ of clusters. With the only exception of the MMH that does not identify the presence of groups ($\hat{G}=1$), all the other models agree in recognizing the existence of two clusters. 
However, 
%by comparing the alternative optimal mixtures, 
one can note that our MMS is the model associated with the best fitting performance (BIC=2144.5). 
The corresponding cluster-specific parameter estimates (mixing weights, concentration and modal orderings) are displayed below.
\begin{center}
\bigskip
$\hat\omega_1=0.61\qquad\hat\theta_1=0.039$\\
\framebox{$\hat{\boldsymbol{\rho}}_1=${ \scriptsize
\includegraphics[scale=.08]{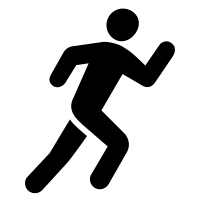}\quad
\includegraphics[scale=.08]{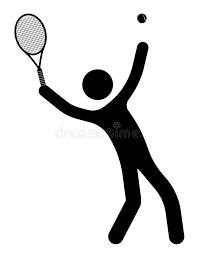}\quad
\includegraphics[scale=.08]{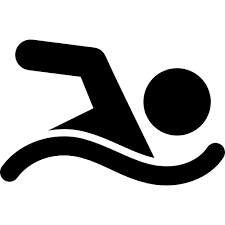}\quad
\includegraphics[scale=.08]{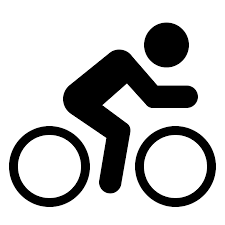}\quad
\includegraphics[scale=.08]{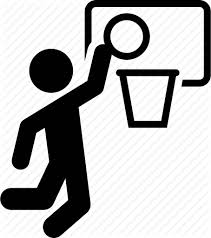}\quad
\includegraphics[scale=.08]{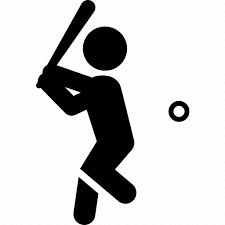}\quad
\includegraphics[scale=.08]{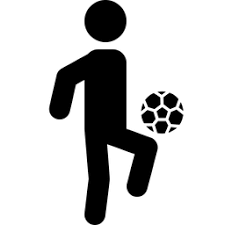}
}
}\bigskip
\end{center}
\begin{center}
$\hat\omega_2=0.39\qquad\hat\theta_2=0.118$\\
\framebox{
$\hat{\boldsymbol{\rho}}_2=${ \scriptsize
\includegraphics[scale=.08]{Baseball.png}\quad
\includegraphics[scale=.08]{Football.png}\quad
\includegraphics[scale=.08]{Basket.png}\quad
\includegraphics[scale=.08]{Cycling.png}\quad
\includegraphics[scale=.08]{Tennis.png}\quad
\includegraphics[scale=.08]{Swimming.png}\quad
\includegraphics[scale=.08]{Jogging.png}
}
}\bigskip
\end{center}
\normalsize
Both groups share the middle position assigned to Cycling, but they are clearly distinguished by top and bottom ranks. In fact, the largest group ($\hat\omega_1=0.61$) is characterized by the preference towards individual sports, whereas team sports are the most-liked by students of the smallest group ($\hat\omega_2=0.39$).

\begin{figure}[b]
    \centering
    \includegraphics[width=0.45\textwidth]{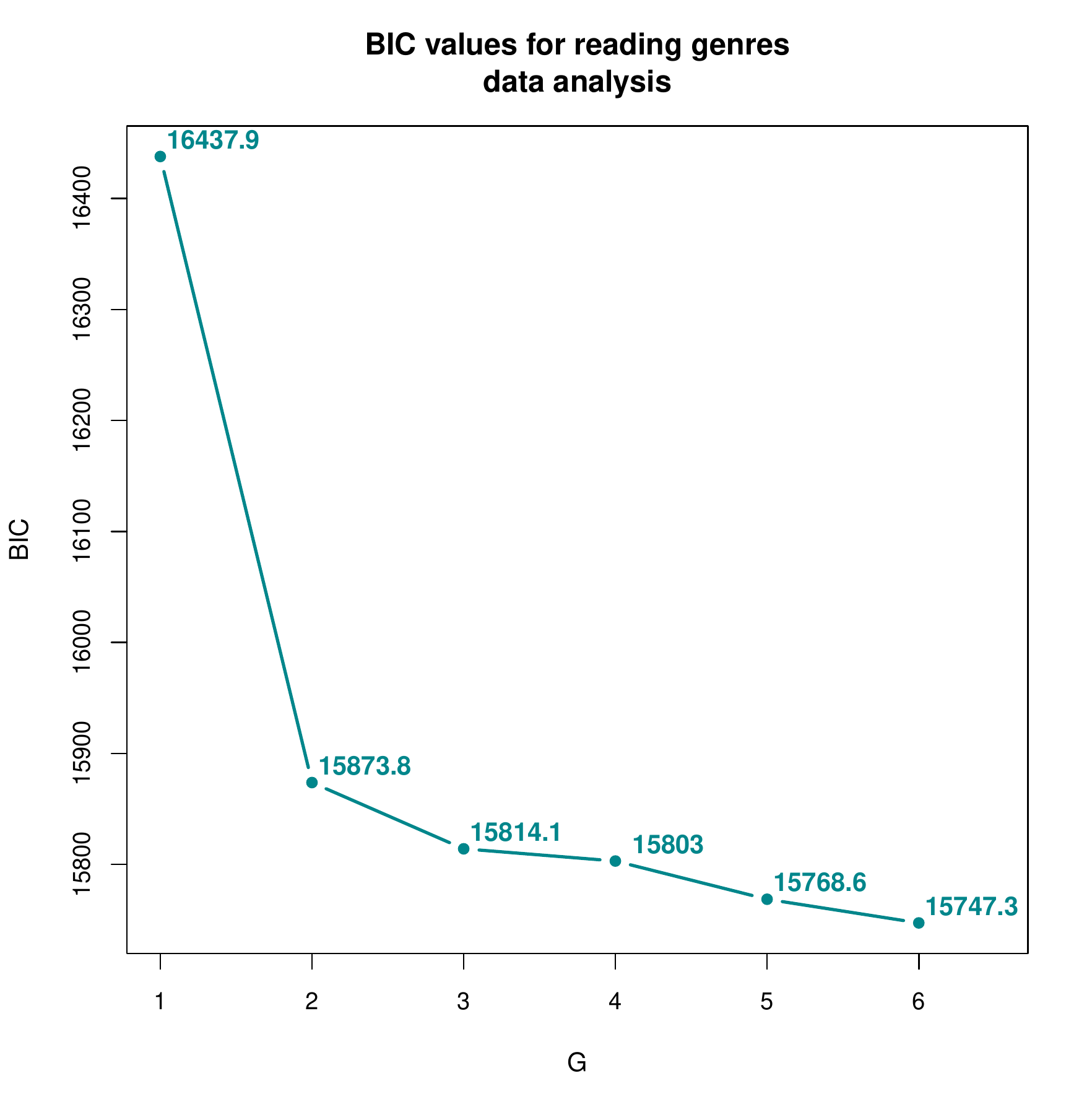}
\caption{Screeplot of BIC values for alternative number of groups in the MMS-mixture fitted to the reading genres data.} 
\label{fig:bicreading}
\end{figure}

\begin{figure}[h!]
    \centering
    \includegraphics[width=0.49\textwidth]{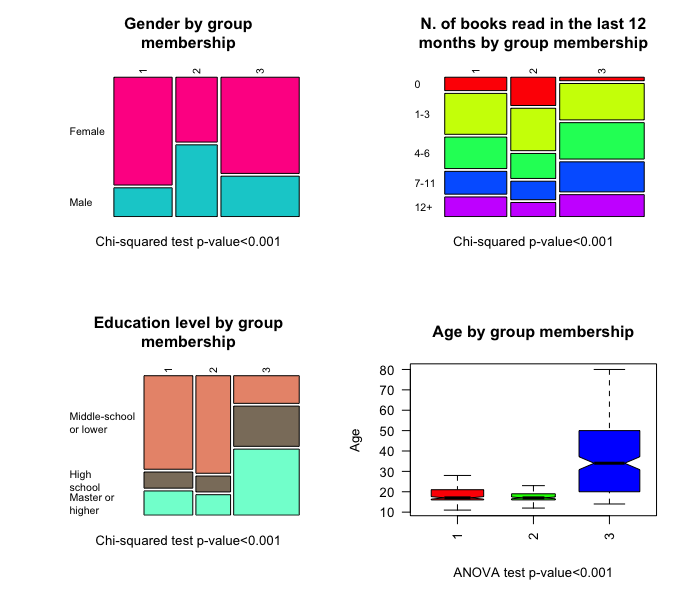}
\caption{Respondents' characteristics statistically associated to the model-based clustering obtained from the application of the MMS-mixture to the reading genres data.} 
\label{fig:reading}
\end{figure}

\begin{table*}[h!]
\centering
\begin{tabular}{lccc}
\hline
Group & $\hat\omega$ & $\hat\theta$ & Consensus\\
\hline
\hline
%1 & 0.42 & 0.048 & (Nov., Cla., Thr., Ess., Bio., Poe., Fan., Com., Hum., Hor., Tee.) \\
%2 & 0.07 & 0.036 & (Fan., Com., Tee., Hum., Cla., Hor., Nov., Thr., Ess., Bio., Poe.) \\
%3 & 0.51 & 0.038 & (Nov., Thr., Fan., Cla., Tee., Hor., Bio., Com., Poe., Ess., Hum.)\\
1 & 0.33 & 0.048 & (Nov., Fan., Cla., Thr., Tee., Poe., Bio., Hor., Com., Hum., Ess.,)   \\
2 & 0.23 & 0.036 & (Fan., Thr., Hor., Com., Nov., Tee., Hum., Cla., Bio., Ess., Poe.,)  \\
3 & 0.44 & 0.054 & (Nov., Cla., Thr., Ess., Bio., Poe., Fan., Com., Hum., Hor., Tee.,) \\
\hline
\end{tabular}
\caption{Parameter estimates of the MMS-mixture fitted to the reading genres data.}
 \label{t:ESTreading}
\end{table*}

\subsection{Analysis of the reading genres}

In this application we illustrate the effectiveness of our method for handling partial rankings. To this purpose, we consider a brand new dataset that has not been analyzed earlier in the ranking literature. This was obtained from an online survey conducted in Italy in 2019 to investigate reading preferences. It was promoted by the municipal administration of Latina (Latium, Italy), in collaboration with Sapienza University of Rome and the School of Government of the University of Tor Vergata. The dataset is composed of $N=507$ partial top-5 rankings of $n=11$ reading genres ordered by the respondents according to their personal preference. The genres are Classic, Novel, Thrillers, Fantasy, Biography, Teenage, Horror, Comics, Poetry, Essay and Humor.

Due to the large sample size and the heterogeneity of the respondents with respect to factors typically impacting on the reading preferences, such as gender, age and education level, we expect the existence of different groups. We explored this hypothesis through the estimation of $G$-component MMS-mixtures with $G=1,\dots,6$. 
Figure \ref{fig:bicreading} shows that BIC values remarkably decrease up to 3 components. Moreover, for $G\geq5$ a cluster with a negligible size is always estimated. According to the elbow rule, we considered $\hat{G}=3$ as the optimal number of groups and reported the corresponding cluster-specific parameter estimates in Table \ref{t:ESTreading}.
%\mc{sicura? io ne avrei scelti 2...} 
The estimates of the modal rankings reveal that more serious and ponderous readings, involving Classic and Essay, are ranked in high positions by respondents of the third cluster, whereas all the genres which are typically the most-liked ones among younger people all occupy bottom positions. This preference pattern is completely reversed in the modal ranking of the second group, with all soft readings (Fantasy, Thrillers, Horror and Comics) ranked in higher positions. The third component represents more hybrid preferences with a mix of the two aforementioned reading types in top and bottom positions. Interestingly, the estimated group structure turned out to be significantly associated to several individual aspects influencing reading attitudes (see Figure \ref{fig:reading}). In fact, the third group is mainly characterized by respondents aged 20 or older, who read many books in the last year and holding a master or doctoral degree. On the other hand, respondents of the second group are mostly teenagers and reported a lower propensity to the reading habit during the last year. Finally, we note that the percentage of males in the second group is remarkably higher than in the other two.

%--------------------------------------------------------------------------

%

% 
%
%\begin{table}[h!]
% \footnotesize
%\centering
%\begin{tabular}{rccc}
% \cline{2-4}
%& Group 1 & Group 2 & Group 3\\
%\cline{2-4}
%$\hat\omega$ & 0.256 & 0.328 & 0.416\\
%$\hat\theta$ & 0.028 & 0.049 & 0.057\\
%\hline
%\textit{Rank 1} & Fantasy & Novel & Novel\\
%\textit{Rank 2} & Thrillers & Classic &   Classic\\
%\textit{Rank 3} & Horror & Essay & Thrillers\\
%\textit{Rank 4} & Novel & Thrillers & Fantasy\\
%\textit{Rank 5} & Teenage & Biography & Poetry\\ 
%\textit{Rank 6} & Comics & Humor & Teenage\\
%\textit{Rank 7} & Humor & Comics &Biography \\ 
%\textit{Rank 8} & Classic & Poetry & Comics\\
%\textit{Rank 9} & Biography & Teenage & Horror\\
%\textit{Rank 10} & Poetry & Fantasy & Essay\\
%\textit{Rank 11} & Essay & Horror & Humor\\
%\hline
%\end{tabular}
%\caption{BIC values of the MMS-mix fitted to the reading genres data.}
% \label{t:ESTreading}
%\end{table}
% 

%\mc{If feasible, providing the respondent characteristics would be helpful. It is better to provide a table of summary statistics of respondents by different clusters instead of a short description at the end of Section 5.2.}

\subsection{An illustration based on gene expression data}\label{ssec:genes}

\begin{table}[t]
\centering
\begin{tabular}{rcccc}
  & \multicolumn{4}{c}{Tumour type}\\
  \cline{2-5}
  Group & NB & BL & RMS & EWS \\ 
  \hline\hline
1 &  12 &   0 &   0 &   0 \\ 
  2 &   0 &  8 &   0 &   0 \\ 
  3 &   0 &   0 &   21 &   1 \\ 
  4 &   0 &   0 &   0 &  22 \\
   \hline
\end{tabular}
\caption{Cross tabulation between the classification resulting from the MMS-mixture application and the true tumour types. }\label{tab:genes}
\end{table}

Finally, we apply our methodology on a micro-array gene expression dataset, to illustrate the ability of our method to analyze heterogeneous rankings of a large number of items. 

We use data adapted from \citet{khan}, which contains micro-array gene expression profiles of four types of small round blue cell tumours (SRBCT) of childhood for $N=64$ patients. These four different cancers, namely neuroblastoma (NB), Burkitt lymphom (BL), rhabdomyosarcoma (RMS), and the Ewing family of tumours (EWS), present a similar histology of SRBCT and thus often leads to misdiagnosis. 

In this illustration we aim at exploiting the MMS-mixture to classify patients into the 4 different cancer types, based on their ranked gene expression profiles.

We select the $n=60$ most differentially expressed genes throughout the sample and, for each patient, we convert the individual quantitative expression profile into a ranking. The rank transformation is sometimes used in genomics applications, because inference based on rankings is both more robust than working on the actual continuous measurements \citep{omics}, and avoids the arbitrary choice of a numerical pre-processing procedure to handle the raw outcomes \citep{mollica14}.

We then estimate a 4-component MMS-mixture and check the agreement between the model-based classification with the one based on the true tumour type. 

Results are reported in Table \ref{tab:genes}, where we see that we classify all tumours correctly except one EWS which is erroneously assigned to the RMS group. 
From Table \ref{tab:genes2}, where we report the classification implied by the simple hierarchical clustering, we note that this method is not able to correctly classify a considerable subset of RMS patients, which are instead erroneously mixed with the BL ones. 

%> fit$mod$bic
%[1] 21537.38
%> fit$mod$alpha
%[1] 0.002892331 0.002817147 0.001527256 0.002433401
%> fit$mod$weights
%[1] 0.1875000 0.1250000 0.3437468 0.3437532

\begin{table}[t]
\centering
\begin{tabular}{rcccc}
  & \multicolumn{4}{c}{Tumour type}\\
  \cline{2-5}
 Group& NB & BL & RMS & EWS \\ 
  \hline\hline
   1 &  12 &   0 &   0 &   0 \\ 
  2 &   0 &  8 &   10 &   0 \\ 
  3 &   0 &  0 &   11 &   1 \\ 
  4 &   0 &   0 &   0 &  22 \\ 
   \hline
\end{tabular}
\caption{Cross tabulation of the hierarchical clustering  groups and true tumour type.}\label{tab:genes2}
\end{table}

\begin{table}[b]
\centering
\begin{tabular}{llll}
  & \multicolumn{3}{c}{Rank}\\
  \cline{2-4}
Group & 1 & 2 & 3 \\ 
  \hline
% $\hat{\boldsymbol{\rho}_1}$ & AF1Q & MAP1B & CRMP1 \\ 
%   $\hat{\boldsymbol{\rho}_2}$ & HLA-DPB1 & MYC & CYFIP2 \\ 
%   $\hat{\boldsymbol{\rho}_3}$ & INS-IGF2 & IGF2 & COL3A1 \\ 
%   $\hat{\boldsymbol{\rho}_4}$ & FCGRT & MYC & OLFM1 \\ 
1 (NB) & AF1Q & MAP1B & CRMP1 \\ 
2 (BL) & HLA-DPB1 & MYC & CYFIP2 \\ 
3 (RMS) & INS-IGF2 & IGF2 & COL3A1 \\ 
4 (EWS)& FCGRT & MYC & OLFM1 \\ 
   \hline
\end{tabular}
\caption{Top-3 genes in the  consensus rankings of the 4 estimated groups. }\label{tab:genes_consensus}
\end{table}

\begin{figure}[t]
    \centering
    \includegraphics[width=0.45\textwidth]{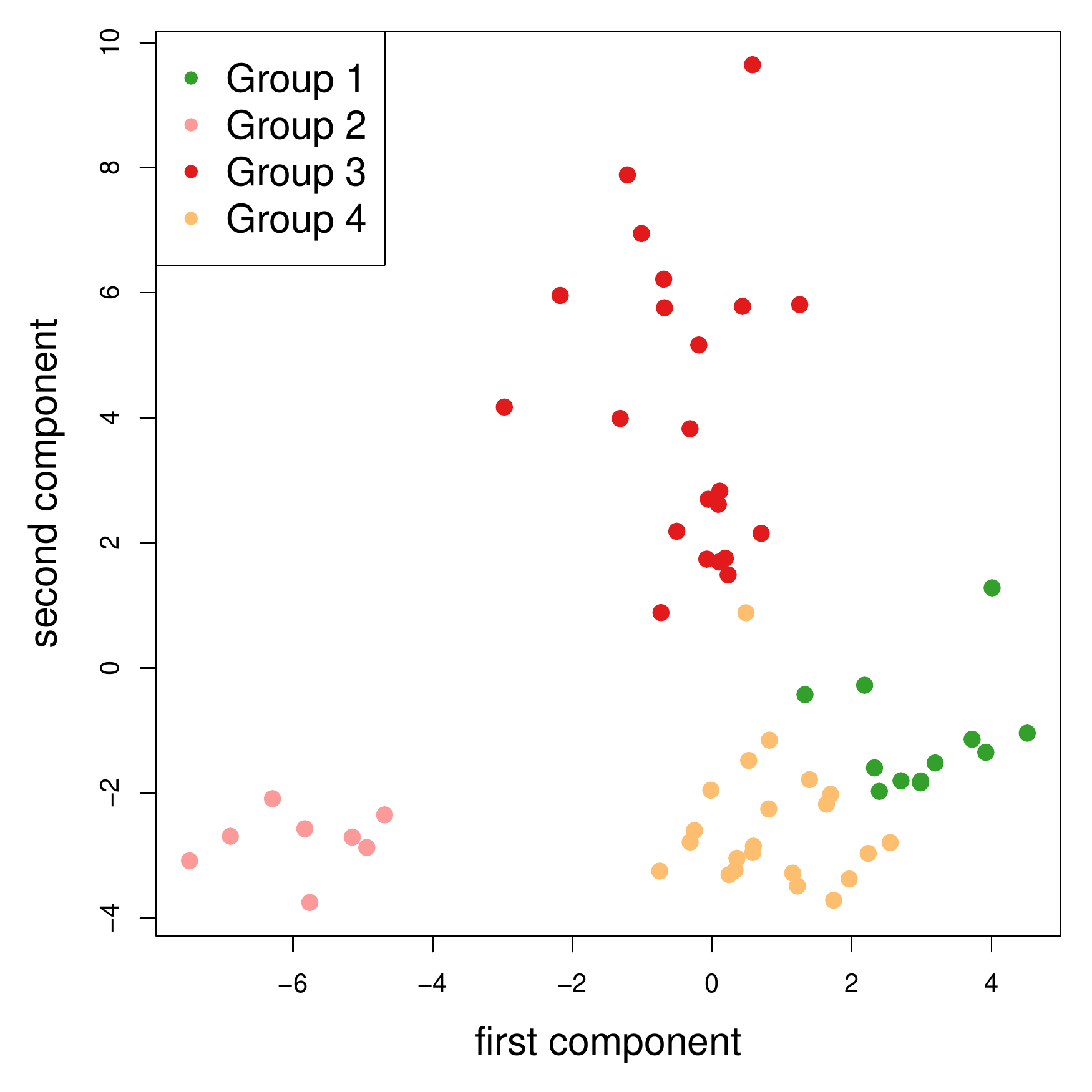}
\caption{Visualization of MMS classification of cancers. The axes are the first two principal components of the original quantitative gene expression levels.} 
\label{fig:cancer}
\end{figure}

In Figure \ref{fig:cancer} we plot the tumours based on their gene expression levels in the 2D space of the top-2 principal components of the gene expression levels of the sample. 
In the figure, the tumours are colored based on their classification obtained with the MMS.
It can be seen that two groups classified by the MMS (groups 2 and 3) are highly separable in this space, while groups 1 and 4 appear less separable.

The mixture of MMS also provides the consensus rankings of the four components, which correspond to the estimated modal orderings of genes for the four cancer types.
%We can therefore characterize 
In Table \ref{tab:genes_consensus}, we report %that
%the different tumours in terms of their 
the top-3 expressed genes of the different tumours, which 
% . For instance, we see that gene AF1Q is top-ranked in group 1 (corresponding to the NB tumour type), HLA-DPB1 in group 2 (corresponding to the BL tumour type), IGF2 in group 3 (corresponding to the RMS tumour type) and FCGRT in group 4 (corresponding to the EWS tumour type). These results 
are in line with the previous literature, that identified them as those capable of diagnosing and discerning the considered cancers types \citep[see, for instance, Table 2 in the work by][]{genetic_top}.

%Table 2 in Diagnosis of the Small Round Blue Cell Tumors Using Multiplex Polymerase Chain Reaction
%https://www.ncbi.nlm.nih.gov/pmc/articles/PMC1867426/

\section{Conclusions and future work}
\label{sec:conc}
Within the class of MMs for the analysis of rankings, the version based on the use of the Spearman distance as metric on the permutation space has been barely explored in the literature, both from an analytical and a computational point of view. One of the main reasons lies in the fact that, unlike the Kendall, the Hamming or the Cayley distances, the Spearman cannot be decomposed into the sum of independent terms, a property which leads to a closed-form solution for the partition function and to convenient simplifications in the estimation procedure. Moreover, with decomposable metrics, more straightforward extensions of the MM to handle partial rankings can be defined. 
Nevertheless, the Spearman distance induces a parametric ranking distribution which parallels the multivariate normal model and actually, similarly to the latter,  %, as indicated by \cite{Marden1995},
%admits a closed-form expression for the location parameter estimate. More precisely, in this work we link a formal result for the uniform ranking model proven in \cite{Marden1995} to the MMS sampling distribution to establish the existence of the 
admits an analytical solution for the MLE of the modal consensus ranking given by the ordering of the average rank vector. %This property is not shared by the MM specifications with other metrics and was not considered earlier in the literature and it . 
Motivated by this theoretical argument and the resulting computational simplifications, which instead do not hold for the other metrics, we extend the MMS to the finite mixture context and develop an efficient and accurate EM algorithm to perform inference on samples of both full and partial rankings drawn from heterogeneous populations and affected by censoring of arbitrary positions. 

Additionally, in order to address the inferential issues arising with a large number of ranked items, this work introduces an approximation of the model normalizing constant and assesses its performance with respect to alternative methods. Finally, the utility of our approach
%to learn patterns of preferences 
has been proven through applications to both synthetic and real-world data.

Our work provides original research that can be further extended in several directions. From a computational perspective, an \texttt{R} package to promote a wider use of the
%proposed methods and to fill the gap of the MLE of the 
MMS and mixtures thereof is currently under construction. From a methodological perspective, model-based clustering could be enhanced by integrating the finite mixture framework with the regression analysis, for identifying significant factors affecting the preferences and better characterizing the groups of rankers. This idea is currently under study 
\citep{moemms} by developing a mixture of experts model for the MMS, %\mc{qui un draft c'è, vogliamo citarlo? oppure semplicemente dire che ci stiamo già lavorando?}, 
in the spirit of \cite{gormley2008mixture}. 
%Finally, a Bayesian version of the present mixture of MMS could be also explored and compared with the approach proposed by \citep{vitelli18}. 

%\bibliographystyle{chicago}

%\bibliography{references}

\begin{thebibliography}{}

\bibitem[\protect\citeauthoryear{Ali and Meil{\u{a}}}{Ali and
  Meil{\u{a}}}{2012}]{Ali2012}
Ali, A. and M.~Meil{\u{a}} (2012).
\newblock Experiments with {K}emeny ranking: What works when?
\newblock {\em Mathematical Social Sciences\/}~{\em 64\/}(1), 28--40.

\bibitem[\protect\citeauthoryear{Alvo and Yu}{Alvo and Yu}{2014}]{AlvoYu2014}
Alvo, M. and P.~L.~H. Yu (2014).
\newblock {\em Statistical Methods for Ranking Data}.
\newblock Frontiers in Probability and the Statistical Sciences. New York, USA:
  Springer.

\bibitem[\protect\citeauthoryear{Andrieu and Roberts}{Andrieu and
  Roberts}{2009}]{Andrieu2009}
Andrieu, C. and G.~O. Roberts (2009).
\newblock The pseudo-marginal approach for efficient {M}onte {C}arlo
  computations.
\newblock {\em The Annals of Statistics\/}~{\em 37\/}(2), 697--725.

\bibitem[\protect\citeauthoryear{Bartholdi, Tovey, and Trick}{Bartholdi
  et~al.}{1989}]{bartholdi1989computational}
Bartholdi, J.~J., C.~A. Tovey, and M.~A. Trick (1989).
\newblock The computational difficulty of manipulating an election.
\newblock {\em Social Choice and Welfare\/}~{\em 6\/}(3), 227--241.

\bibitem[\protect\citeauthoryear{Beckett}{Beckett}{1993}]{beckett}
Beckett, L.~A. (1993).
\newblock {Maximum Likelihood Estimation in Mallows's Model Using Partially
  Ranked Data}.
\newblock In M.~A. Fligner and J.~S. Verducci (Eds.), {\em Probability Models
  and Statistical Analyses for Ranking Data}, New York, pp.\  92--107. Springer
  New York.

\bibitem[\protect\citeauthoryear{Busse, Orbanz, and Buhmann}{Busse
  et~al.}{2007}]{BusseEtal2007}
Busse, L.~M., P.~Orbanz, and J.~M. Buhmann (2007).
\newblock Cluster analysis of heterogeneous rank data.
\newblock In {\em Proceedings of the 24th International Conference on Machine
  Learning}, ICML '07, New York, USA, pp.\  113--120. ACM.

\bibitem[\protect\citeauthoryear{Caron, Teh, and Murphy}{Caron
  et~al.}{2014}]{Caron2014}
Caron, F., Y.~W. Teh, and T.~B. Murphy (2014).
\newblock Bayesian nonparametric {P}lackett-{L}uce models for the analysis of
  preferences for college degree programmes.
\newblock {\em The Annals of Applied Statistics\/}~{\em 8\/}(2), 1145--1181.

\bibitem[\protect\citeauthoryear{Chen, Vansant, Oades, Pickering, Wei, Song,
  Monforte, and Khan}{Chen et~al.}{2007}]{genetic_top}
Chen, Q.~R., G.~Vansant, K.~Oades, M.~Pickering, J.~S. Wei, Y.~K. Song,
  J.~Monforte, and J.~Khan (2007).
\newblock {Diagnosis of the small round blue cell tumors using multiplex
  polymerase chain reaction}.
\newblock {\em The Journal of molecular diagnostics\/}~{\em 9\/}(1), 80--88.

\bibitem[\protect\citeauthoryear{Cram{\'e}r}{Cram{\'e}r}{1938}]{cramer1938nouveau}
Cram{\'e}r, H. (1938).
\newblock Sur un nouveau th{\'e}oreme-limite de la th{\'e}orie des
  probabilit{\'e}s.
\newblock {\em Actualit{\'e}s scientifiques et industrielles\/}~{\em 736},
  2--23.

\bibitem[\protect\citeauthoryear{Crispino and Antoniano-Villalobos}{Crispino
  and Antoniano-Villalobos}{2022}]{crispino2022informative}
Crispino, M. and I.~Antoniano-Villalobos (2022).
\newblock Informative priors for the consensus ranking in the bayesian mallows
  model.
\newblock {\em To appear, Bayesian Analysis\/}.

\bibitem[\protect\citeauthoryear{Crispino, Modugno, and Mollica}{Crispino
  et~al.}{2022}]{moemms}
Crispino, M., L.~Modugno, and C.~Mollica (2022+).
\newblock {The Mallows model with respondents' covariates}.
\newblock Preliminary draft available upon request.

\bibitem[\protect\citeauthoryear{DeConde, Hawley, Falcon, Clegg, Knudsen, and
  Etzioni}{DeConde et~al.}{2006}]{DeConde2006}
DeConde, R.~P., S.~Hawley, S.~Falcon, N.~Clegg, B.~Knudsen, and R.~Etzioni
  (2006).
\newblock Combining results of microarray experiments: A rank aggregation
  approach.
\newblock {\em Statistical Applications in Genetics and Molecular
  Biology\/}~{\em 5\/}(1).

\bibitem[\protect\citeauthoryear{Dempster, Laird, and Rubin}{Dempster
  et~al.}{1977}]{Demp:Lai:Rub}
Dempster, A.~P., N.~M. Laird, and D.~B. Rubin (1977).
\newblock Maximum likelihood from incomplete data via the {EM} algorithm.
\newblock {\em Journal of the Royal Statistical Society: Series B (Statistical
  Methodology)\/}~{\em 39\/}(1), 1--38.
\newblock with discussion.

\bibitem[\protect\citeauthoryear{Diaconis}{Diaconis}{1988}]{Diaconis1988}
Diaconis, P. (1988).
\newblock {\em Group representations in probability and statistics}, Volume~11
  of {\em Lecture Notes--Monograph Series}.
\newblock Hayward, USA: Institute of Mathematical Statistics.

\bibitem[\protect\citeauthoryear{Eliseussen, Fleischer, and Vitelli}{Eliseussen
  et~al.}{2022}]{omics}
Eliseussen, E., T.~Fleischer, and V.~Vitelli (2022).
\newblock {Rank-based Bayesian variable selection for genome-wide
  transcriptomic analyses}.
\newblock {\em Statistics in Medicine\/}, 1--22.

\bibitem[\protect\citeauthoryear{Ellis}{Ellis}{1984}]{ellis1984large}
Ellis, R.~S. (1984).
\newblock Large deviations for a general class of random vectors.
\newblock {\em The Annals of Probability\/}~{\em 12\/}(1), 1--12.

\bibitem[\protect\citeauthoryear{Feigin and Cohen}{Feigin and
  Cohen}{1978}]{FeiginCohen}
Feigin, P. and A.~Cohen (1978).
\newblock {On a Model for Concordance Between Judges}.
\newblock {\em Journal of the Royal Statistical Society. Series B
  (Methodological)\/}~{\em 40\/}(2), 203--213.

\bibitem[\protect\citeauthoryear{Fligner and Verducci}{Fligner and
  Verducci}{1986}]{Fligner1986}
Fligner, M.~A. and J.~S. Verducci (1986).
\newblock Distance based ranking models.
\newblock {\em Journal of the Royal Statistical Society B\/}~{\em 48\/}(3),
  359--369.

\bibitem[\protect\citeauthoryear{G{\"a}rtner}{G{\"a}rtner}{1977}]{gartner1977large}
G{\"a}rtner, J. (1977).
\newblock On large deviations from the invariant measure.
\newblock {\em Theory of Probability \& Its Applications\/}~{\em 22\/}(1),
  24--39.

\bibitem[\protect\citeauthoryear{Gormley and Murphy}{Gormley and
  Murphy}{2008a}]{gormley2008exploring}
Gormley, I.~C. and T.~B. Murphy (2008a).
\newblock {Exploring voting blocs within the Irish electorate: A mixture
  modeling approach}.
\newblock {\em Journal of the American Statistical Association\/}~{\em
  103\/}(483), 1014--1027.

\bibitem[\protect\citeauthoryear{Gormley and Murphy}{Gormley and
  Murphy}{2008b}]{gormley2008mixture}
Gormley, I.~C. and T.~B. Murphy (2008b).
\newblock A mixture of experts model for rank data with applications in
  election studies.
\newblock {\em The Annals of Applied Statistics\/}~{\em 2\/}(4), 1452--1477.

\bibitem[\protect\citeauthoryear{Henery}{Henery}{1981}]{Henery-Royal}
Henery, R.~J. (1981).
\newblock Permutation probabilities as models for horse races.
\newblock {\em Journal of the Royal Statistical Society: Series B (Statistical
  Methodology)\/}~{\em 43\/}(1), 86--91.

\bibitem[\protect\citeauthoryear{Irurozki}{Irurozki}{2014}]{IrurozkiThesis}
Irurozki, E. (2014).
\newblock {\em Sampling and learning distance-based probability models for
  permutation spaces}.
\newblock Ph.\ D. thesis, Department of Computer Science and Artificial
  Intelligence of the University of the Basque Country.

\bibitem[\protect\citeauthoryear{Irurozki, Calvo, and Lozano}{Irurozki
  et~al.}{2016}]{irurozki2016permallows}
Irurozki, E., B.~Calvo, and A.~Lozano (2016).
\newblock Per{M}allows: An {R} package for {M}allows and generalized {M}allows
  models.
\newblock {\em Journal of Statistical Software\/}~{\em 71\/}(12), 1--30.

\bibitem[\protect\citeauthoryear{Jacques, Grimonprez, and Biernacki}{Jacques
  et~al.}{2014}]{Rankcluster}
Jacques, J., Q.~Grimonprez, and C.~Biernacki (2014).
\newblock Rankcluster: An {R} package for clustering multivariate partial
  rankings.
\newblock {\em The R Journal\/}~{\em 6\/}(1), 101--110.

\bibitem[\protect\citeauthoryear{Kendall}{Kendall}{1970}]{kendall1970rank}
Kendall, M.~G. (1970).
\newblock {\em Rank correlation methods}.
\newblock 4th ed. Griffin London.

\bibitem[\protect\citeauthoryear{Khan, Wei, Ringn{\'e}r, Saal, Ladanyi,
  Westermann, Berthold, Schwab, Antonescu, Peterson, and Meltzer}{Khan
  et~al.}{2001}]{khan}
Khan, J., J.~S. Wei, M.~Ringn{\'e}r, L.~H. Saal, M.~Ladanyi, F.~Westermann,
  F.~Berthold, M.~Schwab, C.~R. Antonescu, C.~Peterson, and P.~S. Meltzer
  (2001).
\newblock Classification and diagnostic prediction of cancers using gene
  expression profiling and artificial neural networks.
\newblock {\em Nature Medicine\/}~{\em 7\/}(6), 673--679.

\bibitem[\protect\citeauthoryear{Lee and Yu}{Lee and
  Yu}{2010}]{lee2010distance}
Lee, P.~H. and P.~L.~H. Yu (2010).
\newblock Distance-based tree models for ranking data.
\newblock {\em Computational Statistics \& Data Analysis\/}~{\em 54\/}(6),
  1672--1682.

\bibitem[\protect\citeauthoryear{Lee and Yu}{Lee and
  Yu}{2012}]{lee2012mixtures}
Lee, P.~H. and P.~L.~H. Yu (2012).
\newblock Mixtures of weighted distance-based models for ranking data with
  applications in political studies.
\newblock {\em Computational Statistics \& Data Analysis\/}~{\em 56\/}(8),
  2486--2500.

\bibitem[\protect\citeauthoryear{Lee and Yu}{Lee and Yu}{2013}]{pmr_R}
Lee, P.~H. and P.~L.~H. Yu (2013).
\newblock {An R package for analyzing and modeling ranking data}.
\newblock {\em BMC Medical Research Methodology\/}~{\em 3\/}(1), 1--11.

\bibitem[\protect\citeauthoryear{Liu, Reiner, Frigessi, and Scheel}{Liu
  et~al.}{2019}]{sylvia}
Liu, Q., A.~Reiner, A.~Frigessi, and I.~Scheel (2019, 08).
\newblock {Diverse personalized recommendations with uncertainty from implicit
  preference data with the Bayesian Mallows model}.
\newblock {\em Knowledge-Based Systems\/}~{\em 186}, 104960.

\bibitem[\protect\citeauthoryear{Mallows}{Mallows}{1957}]{Mallows1957}
Mallows, C.~L. (1957).
\newblock Non-null ranking models. {I}.
\newblock {\em Biometrika\/}~{\em 44\/}(1/2), 114--130.

\bibitem[\protect\citeauthoryear{Marden}{Marden}{1995}]{Marden1995}
Marden, J.~I. (1995).
\newblock {\em Analyzing and Modeling Rank Data}, Volume~64 of {\em Monographs
  on Statistics and Applied Probability}.
\newblock Cambridge, USA: Chapman \& Hall.

\bibitem[\protect\citeauthoryear{McCullagh}{McCullagh}{1993}]{McCullagh1993}
McCullagh, P. (1993).
\newblock Models on spheres and models for permutations.
\newblock In M.~A. Fligner and J.~S. Verducci (Eds.), {\em Probability Models
  and Statistical Analyses for Ranking Data}, New York, pp.\  278--283.
  Springer New York.

\bibitem[\protect\citeauthoryear{Meil\v{a} and Bao}{Meil\v{a} and
  Bao}{2010}]{MeilaBao2010}
Meil\v{a}, M. and L.~Bao (2010).
\newblock An exponential model for infinite rankings.
\newblock {\em Journal of Machine Learning Research\/}~{\em 11}, 3481--3518.

\bibitem[\protect\citeauthoryear{Mollica and Tardella}{Mollica and
  Tardella}{2014}]{mollica14}
Mollica, C. and L.~Tardella (2014).
\newblock Epitope profiling via mixture modeling for ranked data.
\newblock {\em Statistics in Medicine\/}~{\em 33\/}(21), 3738--3758.

\bibitem[\protect\citeauthoryear{Mukherjee}{Mukherjee}{2016}]{mukherjee2016}
Mukherjee, S. (2016).
\newblock Estimation in exponential families on permutations.
\newblock {\em The Annals of Statistics\/}~{\em 44\/}(2), 853--875.

\bibitem[\protect\citeauthoryear{Murphy and Martin}{Murphy and
  Martin}{2003}]{MurphyMartin2003}
Murphy, T.~B. and D.~Martin (2003).
\newblock Mixtures of distance-based models for ranking data.
\newblock {\em Computational Statistics \& Data Analysis\/}~{\em 41\/}(3–4),
  645--655.

\bibitem[\protect\citeauthoryear{Murray, Ghahramani, and MacKay}{Murray
  et~al.}{2012}]{murray2012mcmc}
Murray, I., Z.~Ghahramani, and D.~MacKay (2012).
\newblock {MCMC for doubly-intractable distributions}.
\newblock {\em arXiv preprint arXiv:1206.6848\/}.

\bibitem[\protect\citeauthoryear{Qian and Yu}{Qian and Yu}{2019}]{rankdist}
Qian, Z. and P.~L.~H. Yu (2019).
\newblock Weighted distance-based models for ranking data using the {R} package
  {rankdist}.
\newblock {\em Journal of Statistical Software\/}~{\em 90\/}(5), 1--31.

\bibitem[\protect\citeauthoryear{Schwarz}{Schwarz}{1978}]{Schwarz}
Schwarz, G. (1978).
\newblock Estimating the dimension of a model.
\newblock {\em Ann. Statist.\/}~{\em 6\/}(2), 461--464.

\bibitem[\protect\citeauthoryear{Shi, Wei, Wei, Wang, Liu, and Liu}{Shi
  et~al.}{2021}]{cite-key}
Shi, C., B.~Wei, S.~Wei, W.~Wang, H.~Liu, and J.~Liu (2021).
\newblock A quantitative discriminant method of elbow point for the optimal
  number of clusters in clustering algorithm.
\newblock {\em EURASIP Journal on Wireless Communications and
  Networking\/}~{\em 2021\/}(1), 31.

\bibitem[\protect\citeauthoryear{Sloane}{Sloane}{2017}]{sloane17}
Sloane, N. J.~A. (2017).
\newblock The {E}ncyclopedia of {I}nteger {S}equences.

\bibitem[\protect\citeauthoryear{S{\o}rensen, Crispino, Liu, and
  Vitelli}{S{\o}rensen et~al.}{2020}]{BayesMallows}
S{\o}rensen, {\O}., M.~Crispino, Q.~Liu, and V.~Vitelli (2020).
\newblock {BayesMallows: An R Package for the Bayesian Mallows Model}.
\newblock {\em {The R Journal}\/}~{\em 12\/}(1), 324--342.

\bibitem[\protect\citeauthoryear{Varadhan}{Varadhan}{1966}]{varadhan1966asymptotic}
Varadhan, S.~S. (1966).
\newblock Asymptotic probabilities and differential equations.
\newblock {\em Communications on Pure and Applied Mathematics\/}~{\em 19\/}(3),
  261--286.

\bibitem[\protect\citeauthoryear{Vitelli, S{{\o}}rensen, Crispino, Frigessi,
  and Arjas}{Vitelli et~al.}{2018}]{vitelli18}
Vitelli, V., {\O}.~S{{\o}}rensen, M.~Crispino, A.~Frigessi, and E.~Arjas
  (2018).
\newblock {Probabilistic preference learning with the Mallows rank model}.
\newblock {\em Journal of Machine Learning Research\/}~{\em 18\/}(158), 1--49.

\bibitem[\protect\citeauthoryear{Xu, Alvo, and Yu}{Xu
  et~al.}{2018}]{XuAlvoYu2018}
Xu, H., M.~Alvo, and P.~L.~H. Yu (2018).
\newblock Angle-based models for ranking data.
\newblock {\em Computational Statistics \& Data Analysis\/}~{\em 121},
  113--136.

\bibitem[\protect\citeauthoryear{Zhao and Hautam{\"a}ki}{Zhao and
  Hautam{\"a}ki}{2008}]{Zhao08}
Zhao, Q. and V.~Hautam{\"a}ki (2008).
\newblock {Knee Point Detection in BIC for Detecting the Number of Clusters}.
\newblock In {\em Blanc-Talon, J., Bourennane, S., Philips, W., Popescu, D.,
  Scheunders, P. (eds) Advanced Concepts for Intelligent Vision Systems. ACIVS
  2008. Lecture Notes in Computer Science, vol 5259. Springer, Berlin,
  Heidelberg}, pp.\  664--673.

\end{thebibliography}

%\newpage

\bigskip
\begin{center}
{\large\bf SUPPLEMENTARY MATERIAL}
\end{center}

\section*{A. Formal derivation of equation \eqref{approssimazione}}
\label{secA}
%\cm{capire come mai nelle appendici i ranking non compaiono in bold}

Here we give the details for the derivation of equation \eqref{approssimazione}. We first define the rescaled Spearman distance as
$$x_n 
% = \frac{d_S(\boldsymbol{\sigma},\boldsymbol{e})}{2\binom{n+1}{3}}
= \frac{d_n}{d_{\max,n}}
=\frac{c_n-\boldsymbol{e}^T\boldsymbol{\sigma}}{\binom{n+1}{3}} \qquad \boldsymbol\sigma\in \mathcal{P}_n.
$$
For $n\to \infty$, we have that $x_n$ will be densely distributed in the unit interval, $x\defeq \lim_{n\to \infty} x_n \in \left[0,1\right]$.
The result of G{\"a}rtner and Ellis \citep{gartner1977large,ellis1984large} allows us to write the rate function as follows 
\begin{equation}
\label{rate_def}
    \xi\left( x \right) = r\left( x \right) + o (1),
\end{equation} 
with
\begin{equation}
    r\left( x \right) \defeq \lim_{n\to \infty} \frac{1}{n} \log\mathbb{P}\left( x_n \right)
\end{equation}
independent of $n$ and $o(1)$ going to zero when $n\to \infty$.
From the quadratic nature of the Spearman distance, we can derive the property
\begin{equation*}
    d(\boldsymbol{\sigma},\boldsymbol{e}) + d(\boldsymbol{\sigma},\bar{\boldsymbol{e}}) = d_{\text{max},n}
\end{equation*}
with 
%$\boldsymbol{e}$ the identity permutation $\left( 1,2,\dots,n \right)$ and 
$\bar{\boldsymbol{e}}=\left( n,n-1,\dots,1 \right)=(n+1)-\boldsymbol{e}$.
In addition, we can exploit the invariance property of the Spearman distance to obtain
\begin{equation*}
    d(\boldsymbol{\sigma},\bar{\boldsymbol{e}}) = d\left(\left( \boldsymbol{\sigma}\circ\bar{\boldsymbol{e}} \right),\boldsymbol{e}\right),
\end{equation*}

\noindent where the symbol $\circ$ indicates the composition of permutations. 
Given the last two identities and the invertibility of the mapping $\boldsymbol{\sigma}\circ\bar{\boldsymbol{e}}=(\sigma_n,\sigma_{n-1},\dots,\sigma_1)$ %(\bar{e}_{r_1},\dots,\bar{e}_{r_n})$
, we conclude that to every permutation $\boldsymbol{\sigma}$ at distance ${d}_n$ from $\boldsymbol{e}$, it corresponds one permutation at distance $d_{\text{max},n} - {d}_n$ from $\boldsymbol{e}$. This implies $N_{{d}_n} = N_{\left(d_{\text{max},n} - {d}_n\right)}$, and similarly for $r(x)$ we have
\begin{equation*}
    r(x) = r(1-x).
\end{equation*}
The distribution $\mathbb{P}\left( x_n \right)$ is asymptotically unimodal and symmetric, hence its maximum is attained for $x = \frac{1}{2}$. By definition of the rate function \eqref{rate_def} this implies $r\left(\frac{1}{2}\right) = 0$.
% The average value of the normalized distance is $x = \frac{1}{2}$ and, given the asymptotic convergence to a normal distribution, we know that asymptotically the distribution is unimodal; being it also symmetric, in $x=\frac{1}{2}$ is attained the maximum of the distribution, which by the definition of rate function of eq. \eqref{rate_def} implies $r\left(\frac{1}{2}\right) = 0$. 
For the tail values $x\to 0$ and $x\to 1$, we make the assumption of a logarithmic divergence. By considering also the symmetry of $r(x)$, we obtain the tail behaviour
\begin{equation*}
r(x)\sim \alpha_1 \left[\log(x) + \log(1-x)\right] \quad x\to 0^+, 1^-. 
\end{equation*}
For the bulk of the distribution, we use a normal approximation
\begin{equation}
\label{secord_approx}
    r(x) \approx \alpha_1 \left[\log(x) + \log(1-x)\right] + \alpha_2 \, x\,(1-x). 
\end{equation}
We can derive the value of $\alpha_1$ from the knowledge of the values of $N_{{d}_n}$ for ${d}_n=2$,  which give  $x=\binom{n+1}{3}^{-1}\approx \frac{1}{n^3}$. On the other hand,
there is a fraction of $\frac{n-1}{n!}$ permutations at distance $d_n=2$, which implies
\[
r(n^{-3}) = \log(n) + o(1) 
\]
or, equivalently, $\alpha_1=\frac{1}{3}$. By exploiting the fact that $r\left(\frac{1}{2}\right)=0$ and equation \eqref{secord_approx}, we derive the further constraint $8 \log(2) \alpha_1 - \alpha_2 = 0 $, so that only the sub-leading corrections of order $o(1)$ in equation \eqref{rate_def} remain to be estimated.
To perform this estimation, we exploit the known values of $N_{{d}_n}$ for small $n$.
First, we rewrite $\xi(x)$ as
\begin{align*}
    \xi(x) = & r(x)  +  \beta_0 +  \\
& + \beta_1 \left[\log(x) + \log(1-x)\right] +  \beta_2 x\,(1-x)
\end{align*}
with $\beta_0, \beta_1, \beta_2 = o(1)$.
Then, we proceed to estimate $\beta_0, \beta_1, \beta_2$ with a generalized least squares fit. In particular, we select $\beta_0, \beta_1, \beta_2$ in order to minimize the cost function
\begin{equation*}
%    \mathcal{C}\left(\beta_0, \beta_1, \beta_2\right) = 
\sum_{n=n_{\text{min}}}^{n_{\text{max}}}n^2 \sum_{d_n} \left[ \xi\left(\frac{d_n}{d_{\text{max}, n}}\right) - \frac{1}{n}\log\left(\frac{N_{d_n}}{n!}\right) \right]^2,
\end{equation*}
where $n_{\text{min}}$ and $n_{\text{max}}$ are parameters chosen on the basis of the known values of $N_{d_n}$ in the OEIS sequence A175929 \citep{sloane17}. These values are available only for $n \leq 14$, and the first $n$ for which there are no null values of $N_{d_n}$ in the bulk of the distribution is $n=4$. Thus we select the values $n_{\text{min}} = 4$ and $n_{\text{max}}=14$. In addition, we modify the least squares minimization with the $n^2$ prefactor to assign more weight to errors in the estimation for high values of $n$, both because for these values we are closer to the normal limit distribution, and because ultimately we are interested in extending the estimate to values of $n$ greater than the ones for which the $N_{d_n}$ series are available.

Finally, we obtain
\begin{equation*}
    \xi(x) = a_0 +  a_1 \left[\log(x) + \log(1-x)\right] +  a_2\, x\,(1-x),
\end{equation*}
with
%\cm{mettere valori alla quarta cifra decimale come nel codice}
\begin{align*}
%\begin{split}
a_0 =& \alpha_0 \approx - \frac{0.24}{\sqrt{n}},\\
%\qquad 
a_1 =& \alpha_1 + \beta_1 \approx \frac{1}{3}-\frac{0.1784}{\sqrt{n}},\\
%\qquad 
a_2 =& \alpha_2 + \beta_2 \approx \frac{8}{3}\log(2)- \frac{5.5241}{\sqrt{n}}.
%\end{split}
\end{align*}

The highest value of $n$ for which the series $N_{d_n}$ is available is $n=14$. For this value we obtain a root mean square error equal to
$$
%\delta N_{tot, n=14} = 
\sqrt{\frac{\sum_{d_n} \left[ \xi\left(\frac{d_n}{d_{\text{max}, n}}\right) - \frac{1}{n}\log\left(\frac{N_{d_n}}{n!}\right) \right]^2}{\binom{n+1}{3}}} = 0.0027.
$$ 

\section*{B. Derivation of the approximation $\hat{Z}_\text{vMF}(\theta)$}
\label{secB}

Let $n\in\mathbb{N}^+$, $a_n=(n+1)/2$, $b_n=n(n^2-1)/12$ and $c_n=n(n+1)(2n+1)/6
$. Let us denote with $\boldsymbol{r}$
%and $\boldsymbol{\rho}$ two 
a generic rankings of $n$ items, with $\boldsymbol{e}=(1,\dots,n)$ the identity permutation and with $\boldsymbol{y}$ and $\boldsymbol{\pi}$ their standardized versions, i.e. $\boldsymbol{y}=(\boldsymbol{r}-a_n)/\sqrt{b_n}$ and $\boldsymbol{\pi}=(\boldsymbol{e}-a_n)/\sqrt{b_n}$. Let $\mathcal{P}_n$ be the set of all $n!$ rankings and $\mathcal{S}_n$ the set of all $n!$ standardized rankings. 
%
%The normalizing constant of the MMS can be written as
%\begin{equation*}
% \begin{split}   Z(\theta)  = & \sum_{\boldsymbol{r}\in\mathcal{P}_n} \exp\bigg\{-\theta \sum_{i=1}^n (r_i-i)^2\bigg\}=\sum_{\boldsymbol{r}\in\mathcal{P}_n} \exp\bigg\{-\theta \bigg[ 2c_n-2\sum_{i=1}^n r_ii\bigg]\bigg\}=\\
% =&\sum_{\boldsymbol{r}\in\mathcal{P}_n} \exp\{-\theta[ 2c_n-2\boldsymbol r^T\boldsymbol{e}] \}.
% \end{split}
%\end{equation*}
%
We want to show that, by relying on in the result in \citet{XuAlvoYu2018}, the normalizing constant of the MMS 
%expressed as a function of the original ranking $\boldsymbol{r}\in\mathcal{P}_n$, 
over $\mathcal{P}_n$ can be written as 
$$Z(\theta)= e^{-\kappa} C(\kappa,\theta)^{-1}$$
where $\kappa = 2\theta b_n$ and $$C(\kappa,\theta)^{-1}=%\sum_{\boldsymbol{y}\in\mathcal{S}_n} \exp\left\{\kappa \sum_{i=1}^n y_i\pi_i\right\}=
\sum_{\boldsymbol{y}\in\mathcal{S}_n} \exp\big\{\kappa  \boldsymbol y^T\boldsymbol\pi\big\}$$ 
is the normalizing constant of the MMS
%expressed as a function of the standardized ranking $\boldsymbol{y}\in\mathcal{S}_n$
over $\mathcal{S}_n$. Note that
$$\boldsymbol r^T\boldsymbol{e} = b_n\boldsymbol y^T\boldsymbol\pi+n a_n^2\qquad \text{and} \qquad b_n=c_n-n a_n^2.$$
Then, it follows 
\begin{equation*}
 \begin{split}   Z(\theta)  = & \sum_{\boldsymbol{r}\in\mathcal{P}_n} \exp\{-\theta[ 2c_n-2\boldsymbol r^T\boldsymbol{e}] \}\\
 =& \sum_{\boldsymbol{y}\in\mathcal{S}_n} \exp\{-\theta[ 2c_n-2b_n\boldsymbol y^T\boldsymbol\pi-2n a_n^2 \}  \\
  = & \sum_{\boldsymbol{y}\in\mathcal{S}_n} \exp\{-\theta[ 2 b_n-2b_n\boldsymbol y^T\boldsymbol\pi] \} \\
  =& \sum_{\boldsymbol{y}\in\mathcal{S}_n} \exp\{-\kappa+\kappa\boldsymbol y^T\boldsymbol\pi \} = e^{-\kappa} C(\kappa,\theta)^{-1}.
 \end{split}
\end{equation*}
Since \citet[Appendix A]{XuAlvoYu2018} proposed the approximation 
$$C(\kappa,\theta)^{-1}\approx \frac{2^\frac{n-3}{2} n! I_\frac{n-3}{2}(\kappa)\Gamma(\frac{n-1}{2})}{\kappa^\frac{n-3}{2}},$$
we can substitute it in the last expression of $Z(\theta)$, yielding %to the normalizing constant approximation below
\begin{equation}
\label{e:Zvmf}
Z(\theta)\approx \hat{Z}_\text{vMF}(\theta)=
\frac{2^\frac{n-3}{2} n! I_\frac{n-3}{2}(\kappa)\Gamma(\frac{n-1}{2})}{\kappa^\frac{n-3}{2}e^{\kappa}}.
\end{equation}
Note that 
%is slightly different from the one we use here and 
in our manuscript
%In particular, for us 
$\kappa=\theta n(n^2-1)/6=2\theta b_n$, while in \citet{XuAlvoYu2018}, page 114 $\kappa=\lambda n(n^2-1)/12$. Hence, one can recover the exact formula in \citet{XuAlvoYu2018} by simply setting $\theta = \lambda/2$.

\begin{table*}[t]
\centering
\footnotesize
%\small
\subcaptionbox{$n=6$}{
 \begin{tabular}{lrrr}
  \hline
 $\theta$ & $\bar\eta_{\hat\theta_\text{OEIS}}$ & $\bar\eta_{\hat\theta_\text{new}}$ & $\bar\eta_{\hat\theta_\text{vMF}}$ \\
  \hline  \hline
1e-04 & .00387 & .00384 & .00409 \\ 
  5e-04 & .00349 & .00345 & .00369 \\ 
  .001 & .00309 & .00305 & .00329 \\ 
  .005 & .00159 & .00156 & .00178 \\ 
  .01 & .00173 & .00170 & .00183 \\ 
  .05 & .00209 & .00211 & .00215 \\ 
  .1 & .00217 & .00225 & .00369 \\ 
  .5 & .00885 & .00958 & .16272 \\ 
  1 & .01765 & .01776 & .84306 \\ 
   \hline
\end{tabular}}
\hspace{0.3cm}
\subcaptionbox{$n=10$}{
 \begin{tabular}{lrrr}
  \hline
$\theta$ & $\bar\eta_{\hat\theta_\text{OEIS}}$ & $\bar\eta_{\hat\theta_\text{new}}$ & $\bar\eta_{\hat\theta_\text{vMF}}$ \\  \hline  \hline

1e-04 & .00151 & .00148 & .00158 \\ 
  5e-04 & .00115 & .00112 & .00123 \\ 
  .001 & .00083 & .00080 & .00091 \\ 
  .005 & .00042 & .00041 & .00045 \\ 
  .01 & .00045 & .00047 & .00047 \\ 
  .05 & .00088 & .00140 & .00386 \\ 
  .1 & .00126 & .00366 & .01397 \\ 
  .5 & .00662 & .02223 & .23848 \\ 
  1 & .01347 & .01590 & 1.06776 \\
   \hline
\end{tabular}
}\hspace{0.3cm}
\subcaptionbox{$n=14$}{
 \begin{tabular}{lrrr}
  \hline  
$\theta$ & $\bar\eta_{\hat\theta_\text{OEIS}}$ & $\bar\eta_{\hat\theta_\text{new}}$ & $\bar\eta_{\hat\theta_\text{vMF}}$ \\  \hline  \hline
1e-04 & .00115 & .00115 & .00119 \\ 
  5e-04 & .00080 & .00080 & .00085 \\ 
  .001 & .00057 & .00057 & .00062 \\ 
  .005 & .00032 & .00032 & .00034 \\ 
  .01 & .00034 & .00035 & .00040 \\ 
  .05 & .00060 & .00075 & .00686 \\ 
  .1 & .00109 & .00108 & .01912 \\ 
  .5 & .00479 & .01076 & .26687 \\ 
  1 & .01199 & .01534 & 1.15766 \\  
   \hline
\end{tabular}}
\hspace{.5cm}
\subcaptionbox{$n=16$}{
\begin{tabular}{lrr}
  \hline
$\theta$ & $\bar\eta_{\hat\theta_\text{new}}$ & $\bar\eta_{\hat\theta_\text{vMF}}$ \\  \hline
  \hline
1e-04 & .00083 & .00084 \\ 
  5e-04 & .00056 & .00058 \\ 
  .001 & .00032 & .00034 \\ 
  .005 & .00025 & .00025 \\ 
  .01 & .00030 & .00039 \\ 
  .05 & .00058 & .00776 \\ 
  .1 & .00248 & .02001 \\ 
  .5 & .00635 & .27603 \\ 
  1 & .01248 & 1.17938 \\  
   \hline
\end{tabular}
}
\hspace{.5cm}
\subcaptionbox{$n=20$}{
\begin{tabular}{lrr}
  \hline
$\theta$ & $\bar\eta_{\hat\theta_\text{new}}$ & $\bar\eta_{\hat\theta_\text{vMF}}$ \\  \hline\hline
1e-04 & .00051 & .00051 \\ 
  5e-04 & .00030 & .00028 \\ 
  .001 & .00022 & .00019 \\ 
  .005 & .00028 & .00018 \\ 
  .01 & .00041 & .00063 \\ 
  .05 & .00265 & .00842 \\ 
  .1 & .00650 & .02111 \\ 
  .5 & .03650 & .28727 \\ 
  1 & .03655 & 1.21505 \\
   \hline
\end{tabular}}
\hspace{.5cm}
\subcaptionbox{$n=24$}{
\begin{tabular}{lrr}
  \hline
$\theta$ & $\bar\eta_{\hat\theta_\text{new}}$ & $\bar\eta_{\hat\theta_\text{vMF}}$ \\  \hline\hline 
  
1e-04 & .00042 & .00040 \\ 
  5e-04 & .00023 & .00019 \\ 
  .001 & .00019 & .00013 \\ 
  .005 & .00036 & .00023 \\ 
  .01 & .00039 & .00097 \\ 
  .05 & .00439 & .00878 \\ 
  .1 & .00938 & .02168 \\ 
  .5 & .05099 & .29718 \\ 
  1 & .11892 & 1.24676 \\
   \hline
\end{tabular}}

\caption{Average absolute error $\bar\eta_{\hat\theta}$ for the estimation of the concentration parameter associated to inferential procedures based on alternative approximations of the Spearman distance ditribution. Inference was performed on 100 samples simulated from the MMS with varying values of $\theta$ (column 1) and different values of $n\leq 14$ (upper panel) and $n>14$ (lower panel).  %$\hat\theta=\hat\theta_\text{OEIS}$ (column 2), $\hat\theta=\hat\theta_\text{new}$ (column 3) and $\hat\theta=\hat\theta_\text{vMF}$ (column 4). 
}
\label{tab:theta_est1}
\end{table*}

\section*{C. Effect of the approximation on the inference}
\label{secC}

Here we study the effect of the approximation on the inference on $\theta$ by means of simulated data. 
%In fact, the use of the approximate $Z(\theta)$ and of $\text{E}_{\theta}[D_S]$ implies that the estimate of $\theta$ will suffer from some bias. 
We simulate 100 datasets of $N=1000$ rankings from the MMS with $n\in\{6,10,14,16,20,24\}$ and concentration $\theta\in\{0.0001,0.0005,0.001,0.005,0.01,0.05,0.1,0.5,1\}$ and then produce the following three estimates of $\theta$:

\begin{enumerate}
    \item $\hat\theta_\text{OEIS}$,  obtained from the EM procedure, when $Z(\theta)$ and  $\text{E}_{\theta}[D]$ have the exact form (this is available only for $n\leq 14$);
    \item $\hat\theta_\text{new}$, obtained from the EM procedure, when $Z(\theta)$ and  $\text{E}_{\theta}[D]$ are estimated by using our new approximation $\hat N_{d_n}$; 
    \item $\hat\theta_\text{vMF}$, obtained from the procedure described in \cite{XuAlvoYu2018}, that is, a Newton-based recursive approximation solution of \eqref{e:thetaest} when $Z(\theta)$ is approximated with $\hat Z_\text{vMF}(\theta)$, see \cite{XuAlvoYu2018}, page 116.
\end{enumerate}

In Table \ref{tab:theta_est1}, we report the average absolute error $\bar\eta_{\hat\theta}=\sum_{s=1}^{100}\vert\hat\theta_s-\theta\vert/100$ over the 100 samples to evaluate the performance of the three estimates for $n\leq 14$ (upper panel) and $n>14$ (lower panel). From Table \ref{tab:theta_est1}, we observe that $\bar\eta_{\hat\theta_\text{new}}$
%(third columns) 
is smaller than $\bar\eta_{\hat\theta_\text{vMF}}$
%(fourth columns) 
for all cases $n\leq 14$ considered, and the difference among the two errors is quite small for low values of $\theta$, but it grows fast for increasing values of the concentration parameter. This result, in line with the findings described in Section \ref{sec:compZ}, holds generally true also for larger values of $n$, as one can notice in the lower panel of Table \ref{tab:theta_est1}. However, in this case, we obtained that for very small values of $\theta$, $\bar\eta_{\hat\theta_\text{vMF}}$
%(third columns) 
is sometimes smaller than $\bar\eta_{\hat\theta_\text{new}}$
%(second columns). 
The latter happens for instance when $n=20, 24$, and $\theta\leq 0.005$ (see Table \ref{tab:theta_est1} (e) and (f)).

\end{document}